\documentclass[final,5p,times, twocolumn]{elsarticle}

\usepackage{amsmath,amssymb,amsfonts}
\usepackage{algorithm}
\usepackage [noend]{algpseudocode}
\usepackage{algpseudocode}
\usepackage{subfigure}
\usepackage{booktabs}
\usepackage{graphicx,color}
\usepackage{lscape}
\usepackage{textcomp}
\usepackage{colortbl}
\usepackage{tabularx}
\usepackage{url}
\usepackage{hyperref}
\usepackage{placeins}
\usepackage{array}
\usepackage{float}
\usepackage{multirow}
\usepackage{titlesec}
\usepackage{xcolor}

\titleformat{\paragraph}
{\normalfont\normalsize\itshape}{\theparagraph}{1em}{}
\titlespacing*{\paragraph}
{0pt}{3.25ex plus 1ex minus .2ex}{1.5ex plus .2ex}

\journal{Information and Software Technology}

\begin{document}

\begin{frontmatter}



\title{Community Engagement and the Lifespan of Open-Source Software Projects}


\author{Mohit Kaushik, Kuljit Kaur Chahal} 

\affiliation{organization={Department of Computer Science, Guru Nanak Dev University},
            addressline={}, 
            city={Amritsar},
            postcode={143005}, 
            state={Punjab},
           country={India}}

\begin{abstract}
	\textbf{Context:} Open-source software (OSS) projects depend on community engagement (CE) for longevity. However, CE's quantifiable impact on project dynamics and lifespan is underexplored.
	\textbf{Objectives:} This study defines CE in OSS, identifies key metrics, and evaluates their influence on project dynamics (releases, commits, branches) and lifespan.
	\textbf{Methods:} We analyzed 33,946 GitHub repositories, defining and operationalizing CE with validated per-month metrics (issues, comments, watchers, stargazers). Non-parametric tests and correlations assessed relationships with project dynamics and lifespan across quartiles.
	\textbf{Results:} CE metrics significantly associate with project dynamics, with stronger correlations in highly engaged projects. For lifespan, a complex pattern emerged: per-month CE rates are highest in younger projects, declining with age. Yet, a subset of long-lived projects maintains exceptionally high activity. Initial CE bursts appear crucial for establishment, while sustained high engagement drives extreme longevity. Active issue engagement's influence intensifies with age, but passive attention's declines.
	\textbf{Conclusion:} CE dynamically drives OSS project longevity and development. Our findings establish validated CE metrics and offer deeper insights into how diverse community activity patterns contribute to project longevity.
\end{abstract}



\end{frontmatter}

\vfill 
\noindent\rule{\linewidth}{0.4pt} 

\vspace{1mm} 

\small{
	\textbf{Citation Note:} This is the Accepted Manuscript of an article published in \textit{Information and Software Technology}. The final Version of Record is available via subscription at: \url{https://doi.org/10.1016/j.infsof.2025.107914}
}
\vspace{5mm} 



\section{Introduction}\label{sec:intro}
The concept of Open Source Software (OSS) emerged in the late 1990s, rapidly gaining widespread acceptance. By October 2021, an estimated 90\% of companies reportedly utilized open source~\cite{Octoverse2022}. While OSS development has historically been volunteer-driven, there has been a notable increase in the involvement of large organizations. This includes their contributions to and acquisitions of companies that provide significant products or platforms central to the OSS ecosystem, exemplified by Microsoft's acquisition of GitHub, a widely used platform for hosting software projects, and IBM's acquisition of Red Hat~\cite{IBMACQ}. Many large and active OSS projects continue to be sustained by developers contributing in their own time. The OSS market has experienced substantial growth, valued at over \$17 billion in 2019~\cite{OSSMarket}, expanding to \$25.03 billion within three years, and projected to grow at a compound annual growth rate (CAGR) of 16.9\% from 2023 to 2030~\cite{OSSMSize}. This trend underscores the growing importance and adoption of open-source technologies.

The OSS paradigm is inherently collaborative, relying on diverse communities of developers to create and improve software. The developer community forms the core of OSS, with members' contributions being vital to project sustainability~\cite{Schueller2022, Malgonde2023ResilienceIT, Gamalielsson2014}. However, mere community presence is insufficient; establishing trust and interpersonal relationships is important for fostering engagement, collaboration, innovation, and a sense of belonging within the open-source ecosystem~\cite{CetrilCommunity}.

Case studies across diverse disciplines (including health~\cite{Closson2016}, research~\cite{Yalegama2016}, and urban/rural development~\cite{Bamberger1987}) consistently emphasize community engagement (CE) as a key indicator for project success and sustainable development. While CE's importance is recognized in various contexts, its role in OSS remains largely underexplored. Existing OSS research has focused on community diversity, participation, health, and community building~\cite{Bosu2019, McDonald2013, Lumbard2024, West2005, Hannemann2013}. Understanding how CE contributes to OSS project sustenance necessitates first defining and measuring it within this specific context.

To address this gap, our study defines CE and proposes metrics for quantifying it in OSS projects. It further investigates CE's impact on project lifespan, thereby contributing to a deeper understanding of the dynamics between CE and project outcomes in the OSS ecosystem.

The rest of the paper is organized as follows: Section~\ref{sec:bck} provides foundational understanding of CE. Section~\ref{sec:LR} reviews existing literature on OSS community aspects. Section~\ref{sec:RM} details research methodology for defining, measuring CE, and analyzing its impact on project lifespan. Section~\ref{sec:DMCE} defines and measures CE in OSS, discussing key aspects, proposing metrics, and presenting their initial validation. Section~\ref{sec:CEPD} presents results for CE metrics and project dynamics. Section~\ref{sec:CEPDLS} investigates CE metrics' impact on project lifespan. Section~\ref{sec:discuss} discusses the implications of our findings. Finally, Section~\ref{sec:concl.} concludes the paper and outlines future research.

\section{Background}\label{sec:bck}
 A \textit{Community} is a group of individuals sharing commonalities like geographical proximity, shared interests (values, beliefs), interactions, a sense of belonging, mutual support, and shared processes~\cite{MacQueen2001, Capece2013, Bettez2013, Nieckarz2005, Vogl2017, Rothblum2010, Theodori2005}. CE acts as a catalyst, fostering collaboration among members to achieve common goals and encourage active contribution.

Classifying communities is essential for comprehending CE. Brint~\cite{Brint2001} categorizes them by geographic (e.g., place-based) and choice-based factors (e.g., activity-based, virtual). Additionally, communities are classified by purpose: interest (shared passion), location (place-oriented actions), action (striving for change), or practice (shared concern on a topic)~\cite{Cantador2011, Theodori2013, Zacklad2003, Grout2022}. Different community types may adopt varied approaches to measuring engagement, with distinct key aspects. For instance, employee engagement emphasizes leadership, rewards, and professional growth opportunities~\cite{Ritu2024}.

CE lacks a universal definition, varying across domains, disciplines, and communities~\cite{Natarajarathinam2021}. The Centers for Disease Control and Prevention (CDC) defines CE as a collaborative process where groups share proximity or similar situations to address well-being, acting as a vehicle for positive environmental and behavioral change~\cite{Atlanta1997}. This widely adopted definition highlights collaboration and shared purpose. Other definitions emphasize aspects like collaborative action and learning for common future visions in a multidimensional context~\cite{Born2008}, attracting stakeholder participation in placemaking~\cite{Hes2019}, or actively involving community members in all project aspects to foster collaboration and align with community priorities~\cite{Yalegama2016}.

A review of the literature reveals core CE elements across disciplines. For example, Cheryl et al.~\cite{Kra2019} define it as engaging healthcare stakeholders, while Closson et al.~\cite{Closson2016} describe involvement in interpersonal interactions outside the home. These definitions frequently use similar terminology: collaboration, interaction, participation, involvement, and communication. Exploring these key aspects is crucial, as all definitions hinge upon them, yet their specific manifestation is context-dependent. Employee engagement emphasizes psychological factors and interaction~\cite{Jeyabharathy2023}, while citizen engagement prioritizes participation and commitment~\cite{hovhannisyan2020theoretical}.

While CE is based on collaboration and interaction, the literature presents various engagement approaches. These include traditional methods (e.g., flyers, newsletters) and interactive ones (e.g., workshops, public meetings, websites, surveys)~\cite{Hes2019}. Taffere et al.'s systematic review highlights methods like community involvement, consultation, trust-building, transparency, and incorporating resident perspectives~\cite{Taffere2023}. Additional approaches, such as forums and social media engagement, are also employed.

In conclusion, CE lacks a universal definition due to its context-dependent meaning, with definitions built upon key aspects and characteristics that vary across disciplines. This background, highlighting CE's diverse definitions, approaches, and classifications, provides a foundation for comprehensively defining CE within the specific context of OSS.

\section{Literature Review}\label{sec:LR}
Building on the previous section's discussion of CE, this section focuses on CE within the context of OSS.

Early OSS development, once dominated by ``lone hackers'' or small elite teams, has evolved into large, collaborative communities. These function as Communities of Practice (CoPs), where contributors share interests and practices~\cite{Ye2003}.

Tamburri et al.~\cite{Tamburri2019} introduced `Yielding Open-Source Health Information' (YOSHI) to classify OSS communities using engagement metrics such as pull-request comments, monthly commit activity, active member count, watcher lists, and collaboration distributions across 25 GitHub projects. Daniel et al.~\cite{Daniel2012} examined diversity's impact on CE, finding contribution-based disparity positively correlated with CE, while culture-based separation diversity showed a negative correlation.

Chodapaneedi et al.~\cite{Chodapaneedi2017} analyzed 10 Apache projects using metrics like lines of code, commits, files, hours, response time, and bug comments. Nijsse et al.~\cite{Nijsse2023} identified CE dimensions in blockchain projects using authorship, commits, comments, pull requests, and public interest metrics (stars, forks, contributors), validated via confirmatory factor analysis. Their follow-up study~\cite{Nijsem2023} linked engagement to software health.

While individual developer engagement is well-studied, community-level CE has received limited direct attention~\cite{Tamburri2019, Daniel2012}. Most research emphasizes individual factors like motivation, retention, and contributions~\cite{Ye2003, Stewart2006, Zhou2015, Wang2018, Constantinou2017}. Aggregating these metrics can reveal community-level engagement. Kaur et al.~\cite{Kaur2022}, in a systematic review, emphasized CE's role in sustainability and noted that even indirectly related metrics can be informative.

Several studies offer proxies for CE. Ye et al.~\cite{Ye2003} used email counts and code contributions; Stewart et al.~\cite{Stewart2006} examined team size and effort, touching on trust and communication. Zhou et al.~\cite{Zhou2015} studied newcomer engagement via comments, fixed issues, response time, and user count. Wang et al.~\cite{Wang2018} modeled LTC identification using early behaviors like comments, merged pull requests, followers, and project popularity. Norikane~\cite{Norikane2017} emphasized mentorship and positive interactions. Constantinou et al.~\cite{Constantinou2017} linked retention to communication and commit frequency.

Community dynamics have also been explored through repository mining. GrimoireLab~\cite{duenas2018perceval} provides tools for extracting metrics from GitHub, GitLab, mailing lists, and issue trackers, enabling systematic CE evaluation.

The CHAOSS initiative~\cite{goggins2021making} defines standardized metrics across code development, responsiveness, risk, and DEI, supporting analysis of retention and communication. RISCOSS~\cite{franch2014layered} translates community data into business-aligned indicators, inferring reliability and sustainability from engagement patterns.

Developer experience influences long-term viability. Storey et al.~\cite{noda2023devex} introduced DevEx, highlighting feedback loops, cognitive load, and flow state as drivers of productivity and satisfaction. Positive experiences correlate with consistent contributions and healthier communities.

In summary, while individual participation and motivation in OSS are well-documented, and frameworks like GrimoireLab, CHAOSS, RISCOSS, and DevEx offer valuable insights into project health and contributor support, there remains a need for focused analysis on how quantifiable, aggregated CE metrics influence long-term sustainability. This study addresses that gap.

\section{Research Design Overview}\label{sec:RM}

This section outlines the research design, defining CE in OSS through metrics and measures, and investigating its impact on project lifespan. To achieve this, we formulated three main research questions (RQs) and their sub-questions.

\subsection{Research Questions and Methodologies}\label{subsec:RQ_Methods} 

\begin{enumerate}\label{RQuestions}
	\item \textbf{Define and measure CE in the context of OSS?}\label{RQ1}
	\begin{enumerate}
		\item What is CE within the context of open-source 	software?\label{RQ1a}
		\item What are the various metrics of CE?\label{RQ1b}
	\end{enumerate}
	\textbf{Methodology:} This RQ established a robust definition via comprehensive literature review. Subsequently, exploratory factor analysis (EFA) and cross-validation identified key CE metrics. Finally, face validity and Weyuker properties validated the proposed metrics.
	
	\item \textbf{Which dimensions of CE affect project dynamics?}\label{RQ2}
	\begin{enumerate}
		\item Which specific CE factors have the most significant correlations with various project dynamics?\label{RQ2a}
		\item Does this correlation vary in high or low active engagement level projects?\label{RQ2b}
		\item Which identified factors of CE influence the project dynamics?\label{RQ2c}
		\item Does the effect of CE metrics on project dynamics vary with project age?\label{RQ2d}
	\end{enumerate}
	\textbf{Methodology:} To investigate CE's relationship with per-month project dynamics (e.g., commits, branches, releases), we employed Spearman's rank correlation. Log-transformed data and OLS regression assessed CE's influence, particularly examining age-dependent effects via interaction terms. Bootstrapping evaluated the significance of differences in correlation coefficients between high/low-active engagement level projects.
	
	\item \textbf{Which CE and project dynamic metrics contribute to the lifespan of OSS projects?}\label{RQ3}\\
		\textbf{Methodology:} This RQ examined the connection between per-month CE and project dynamic metrics and project lifespan. Projects were categorized into lifespan quartiles based on their active lifespan. We then used non-parametric statistical tests (Wilcoxon rank-sum test and Cliff’s delta) to compare these per-month rates across different lifespan groups, assessing observed differences and effect sizes.
	
\end{enumerate}

\subsection{Data Collection}\label{subsec:Data}

For our CE analysis in OSS, we used a dataset of 1.32 million repositories from Dabic et al.~\cite{Dabic2021}, originally sourced from GitHub search (\url{https://seart-ghs.si.usi.ch/}). All repositories in the dataset were explicitly checked for the presence of an OSI-approved open-source license (e.g., MIT, Apache 2.0, GPL) to confirm their classification as OSS projects, rather than merely software projects hosted on GitHub. To focus on active software development, this dataset included repositories with a minimum of 101 Lines of Code (LOC). Attributes relevant to CE were selected based on literature, and additional data on pull request and issue comments were collected using the PyGitHub library\footnote{https://pypi.org/project/PyGithub/}. Metrics such as Pull Request Acceptance Rate and Issue Resolution Rate were also calculated.

\subsubsection{Attributes Relevant to CE}
\begin{enumerate}
	\item \textbf{Number of contributors}: Unique individuals who have committed, opened issues, or submitted pull requests~\cite{Tamburri2019,Nijsse2023}.
	\item \textbf{Number of pull requests}: Total pull requests submitted~\cite{Nijsse2023}.
	\item \textbf{Number of issues}: Total issues opened~\cite{Hata2022}.
	\item \textbf{Number of pull request comments}: Total comments on pull requests~\cite{Tamburri2019,Nijsse2023}.
	\item \textbf{Number of issue comments}: Total comments on issues~\cite{Tamburri2019,Nijsse2023}.
	\item \textbf{Number of watchers}: Total users watching the repository for updates~\cite{Tamburri2019}.
	\item \textbf{Pull Request Acceptance Rate}: Percentage of merged or successfully closed pull requests.
	\item \textbf{Issue Resolution Rate}: Percentage of resolved or successfully closed issues.
	\item \textbf{Number of commits}: Total code changes made to the repository~\cite{Tamburri2019,Nijsse2023}.
\end{enumerate}

To further refine the dataset, we applied the following criteria: repositories must 1) have had no commits within the last six months (excluding active projects for lifespan calculation~\cite{liao2019prediction}); 2) not be a fork (to differentiate original project lifespans); 3) have at least three contributors (indicating community involvement beyond personal repositories~\cite{xia2022predicting}); and 4) possess at least one pull request and one issue (reflecting community interaction and enabling Acceptance Rate/Resolution Rate calculations). Attributes like `createdAt` and `lastCommit` were retained for lifespan calculation.

After removing initial null values, 1.03 million repositories remained. These were then subjected to further refinement based on our exclusion criteria:
\begin{itemize}
	\item 743,165 repositories with commits in the last six months were classified as active and excluded.
	\item 25,190 repositories were identified as forks and excluded.
	\item 506,301 had fewer than three contributors and were removed.
	\item 232,043 had zero total issues, and 221,296 had zero total pull requests; these were also excluded.
\end{itemize}

A single repository could satisfy multiple exclusion criteria. After applying all criteria, the final dataset comprised 291,676 repositories from the original 1.32 million. For these, Pull Request Acceptance Rate and Issue Resolution Rate were calculated as ratios of merged pull requests to total pull requests and resolved issues to total issues, respectively.

Despite using an authenticated GitHub REST API, which provides comparatively higher limit than unauthenticated, the sheer volume of data required, particularly for collecting all pull request and issue comments across 291,676 repositories. Therefore, a sample of 33,946 repositories out of the total 291,676 was selected for detailed analysis, with its size determined using a 95\% confidence level, a 0.5 margin of error, and a 50\% population proportion to ensure statistical representativeness of the overall population.

\subsection{Data Preprocessing for Community Engagement Analysis}

To enable fair comparisons of community engagement intensity across projects with varying durations, all accumulated metrics (including commits, watchers, contributors, total issues, open issues, total pull requests, issue comments, pull request comments, forks, and stargazers) were normalized by each repository’s active lifespan. This approach helps mitigate the bias introduced by longer project histories, which naturally yield higher raw counts.

Active lifespan was defined as the period between a repository’s \texttt{createdAt} and \texttt{lastCommit} timestamps, measured in days and converted into months using an average of 30.44 days per month. Projects with a lifespan of zero days were excluded from the analysis as they could not be reliably used for per-month metric calculations. Normalizing cumulative totals into per-month rates aligns with established practices in empirical OSS research~\cite{Nijsse2023}, facilitating meaningful intensity comparisons across projects and stages of development.

\section{Defining and Measuring CE}\label{sec:DMCE}

This section presents the methods, results, and analysis for our first research question.

\subsection{Defining CE}\label{subsec:DCE}

To address RQ~\ref{RQ1a}, we define CE within OSS. Building on existing definitions, contexts, and characteristics, we synthesize an operational definition for OSS, supported by relevant literature.

Our analysis of CE across various contexts, including OSS, reveals: 1) CE lacks a universal definition, adapting to specific purposes (e.g., public health vs. software development); 2) CE is multidimensional (collaboration, involvement, communication); 3) community type influences engagement; and 4) key aspects vary across contexts (e.g., leadership in employee engagement). This forms the foundation for our customized OSS CE definition.

Considering the OSS community as a \emph{Community of Practice} (CoP)~\cite{Ye2003}, we examine CoP key aspects. In workplace communities, these include sustained relationships, shared perspectives, and practices~\cite{Zhang2022}. Other CoP factors include learning through participation, shared resource development, trust-building, common interests, identity, domain expertise, and effective communication fostering relationships, patience, support, and encouragement~\cite{santos2018communities, Langley2017,Grout2022}.

Before defining CE for OSS, we examine collaborative activities and communication channels. OSS is collaboratively developed by users (contributors)~\cite{moradi2021community}. Contributions vary, including code (new features, bug fixing, refactoring) and non-code (documentation, community management)~\cite{Ye2003}. Both technical and non-technical contributors form the OSS community, where effective communication is vital for project sustainability~\cite{Han2023}. Contributors communicate through mailing lists, IRC, discussions, and forums~\cite{Hata2022}. They connect through shared values and goals, collaboration, mutual support, openness, transparency, recognition, and appreciation~\cite{liang2022understanding,Hanna2018,Trinkenreich2020}. Diverse communication channels foster collaboration and knowledge sharing, thereby fostering engagement~\cite{CetrilCommunity}.

Having examined vital collaborative activities and communication channels in OSS communities, we now define CE in open-source development as: ``a continuous process of building and fostering collaborative relationships characterized by mutual respect and shared values through communication channels (e.g., mailing lists, forums, real-time communication tools) among diverse stakeholders (developers, users, contributors, organizations) to build high-quality OSS and foster a sustainable community."

The definition effectively captures key aspects of CE as an ongoing, interactive relationship between a community and its members~\cite{Taffere2023, MacQueen2001}. OSS development fosters a Community of Practice, where diverse members work to achieve shared goals: building high-quality software and fostering a sustainable community~\cite{Ye2003}. OSS communities collaborate by sharing resources on platforms like GitHub and engaging in discussions about bug fixes, feature requests, and code improvements through mailing lists, issues, and pull requests, fostering consistent contributions and strengthening the CoP.

\subsection{Measuring CE}\label{subsec:CEMetric}
Building upon our novel and comprehensive definition of CE in OSS (as detailed in Section~\ref{subsec:DCE}), this section addresses RQ~\ref{RQ1b} by detailing how we operationalized this multifaceted concept through empirically derived and validated metrics. While individual metrics for OSS activity are widely recognized, our primary contribution in this section lies in systematically identifying how a combination of these metrics can robustly represent the distinct dimensions of CE as defined by our framework. Measuring these specific dimensions, which reflect active participation and interaction within the OSS community, is fundamental for fostering sustainable communities and project longevity~\cite{Kaur2022} and enabling rigorous quantitative research. Metrics identified from the literature were collected and analyzed using Python and R, with Factor Analysis specifically conducted in R to reveal their underlying structure.

\subsubsection{Methods}\label{RQ1Methods}

\paragraph{Data Overview}\label{para:DA}

The dataset comprises 33,946 GitHub repositories. Engagement-related attributes were selected based on prior literature and normalized by each repository’s active lifespan to reflect intensity rather than accumulation. All time-based metrics (such as commits, issues, and comments) were converted into per-month rates, to account for variability in project duration. Metrics that represent ratios, such as Pull Request Acceptance Rate (PRAR) and Issue Resolution Rate (IRR), were excluded from lifespan normalization.

The following indicators were used to assess community engagement:

\begin{itemize}
	\item \textbf{Commits per month (CPM)}: Reflects the rate of active development and contributor effort.
	\item \textbf{Issue Comments per month (IC/m)} and \textbf{Pull Request Comments per month (PRC/m)}: Indicate collaboration and discussion intensity within the community.
	\item \textbf{Total Issues (TI/m)}, \textbf{Open Issues (OI/m)}, \textbf{Total Pull Requests (TPR/m)}, and \textbf{Open Pull Requests (OPR/m)}: Signal the frequency of problem reporting and improvement proposals. A high number of unresolved issues or open PRs may suggest bottlenecks or review delays.
	\item \textbf{Contributors per month (CNT/m)}: Represents diversity and attractiveness of the project to contributors.
	\item \textbf{Watchers (WT/m)} and \textbf{Stargazers (STR/m)}: Serve as indicators of user interest and visibility within the community.
	\item \textbf{Forks per month (FK/m)}: Measures interest in creating derivative versions, suggesting vibrant engagement and experimentation.
	\item \textbf{Pull Request Acceptance Rate (PRAR)}: A high rate suggests active maintainer participation and inclusive contribution practices.
	\item \textbf{Issue Resolution Rate (IRR)}: Reflects effectiveness in managing and resolving reported problems, promoting sustained contributor involvement.
\end{itemize}

\paragraph{Factor Analysis}\label{subsubsec:FA}

Factor analysis (FA) is a multivariate statistical method used to identify latent factors explaining relationships among observed variables. It simplifies multidimensional data by identifying a smaller set of common factors. FA relies on key statistical assumptions~\cite{awe2022comprehensive}:

\begin{itemize}
	\item Sample size: Adequate for reliable analysis.
	\item Adequacy of Correlations: Moderate, positive correlations.
	\item No Multicollinearity: Variables should not exhibit high multicollinearity.
	\item Minimal Missing Data: Excessive nulls can affect accuracy.
	\item Level of Measurement: Variables ideally at interval/ratio level.
\end{itemize}

Assuming these conditions are met, the analysis proceeds. Null values were confirmed absent (Section~\ref{sec:RM}) and data adhered to appropriate measurement levels. The analysis proceeded in two steps: 1) Exploratory Factor Analysis (EFA), and 2) Cross-Validation.

\paragraph{Exploratory Factor Analysis}\label{MEFA}

Exploratory Factor Analysis (EFA) is a statistical method that identifies unknown relationships and simplifies complex datasets by reducing dimensions into a smaller set of underlying latent constructs. EFA begins with sample adequacy tests, including the Kaiser-Meyer-Olkin (KMO) measure and Bartlett's test of sphericity, which determine data suitability and assess if variables are uncorrelated~\cite{awe2022comprehensive}. A minimum sample size of 200 observations is commonly suggested for reliability~\cite{awe2022comprehensive}. The number of factors is then determined using methods like eigenvalues, scree plots, and parallel analysis. Factor extraction methods influence interpretation. To address multicollinearity, the Variance Inflation Factor (VIF) is used; a VIF score below 5 indicates no multicollinearity among independent variables~\cite{Kim2019}.

\paragraph{Cross Validation}

The EFA model can be validated using confirmatory factor analysis (CFA) or cross-validation, achieved by splitting data into training and testing sets~\cite{Nijsse2023}. Cross-validation was chosen for our study to assess generalizability to unseen data, avoiding bias and overfitting, which is important for exploring underlying engagement metric structures. This process was performed using the \textit{caTools}\footnote{https://cran.r-project.org/web/packages/caTools/index.html} package in R.

\subsubsection{Results}
\paragraph{Data Distribution}\label{RDD}
Assessing data distribution is essential for factor analysis, as the extraction method's appropriateness depends on underlying distributional characteristics. Our analysis began by examining descriptive statistics (Table~\ref{tab:tab4}) for the per-month normalized attributes. These revealed diverse repository attributes, such as contributors per month ranging from 0.0188 to 44746.8.

Despite normalization by active lifespan, most accumulated attributes, when expressed as per-month rates, still exhibited a high concentration towards lower values (e.g., open pull requests per month 25th and 50th percentiles remain 0), indicating a persistent, highly skewed distribution. This aligns with general observations from GitHub's advanced search (https://github.com/search/advanced), where a large proportion of repositories show low activity rates, suggesting a Pareto-like distribution even for per-month metrics. This trend motivated further distributional analysis to understand the exact nature of this skewness.

\begin{table}[!h]
	\centering
	\caption{Descriptive Statistics of Per-Month Repository Attributes}
	\label{tab:tab4}
	\setlength{\tabcolsep}{2pt}
	\resizebox{\linewidth}{!}{
		\begin{tabular}{l r r r r r r r r r}
			\toprule
			\textbf{Attribute} & \textbf{Mean} & \textbf{Std Dev} & \textbf{Min} & \textbf{25\%} & \textbf{50\%} & \textbf{75\%} & \textbf{Max} \\
			\midrule
			CPM & 507.9813 & 64996.9200 & 0.0994 & 1.4353 & 3.5268 & 10.4126 & 11794250.0000 \\
			WT/m & 1.0952 & 13.8175 & 0.0000 & 0.0949 & 0.2133 & 0.5205 & 1248.0400 \\
			CNT/m & 2.7896 & 246.0340 & 0.0188 & 0.0971 & 0.1746 & 0.3571 & 44746.8000 \\
			TI/m & 1.9165 & 12.0952 & 0.0068 & 0.1320 & 0.3667 & 1.1211 & 1430.6800 \\
			OI/m & 0.6610 & 7.8545 & 0.0000 & 0.0201 & 0.0925 & 0.3158 & 1065.4000 \\
			TPR/m & 2.8993 & 48.5304 & 0.0063 & 0.1620 & 0.4174 & 1.3135 & 6070.3768 \\
			OPR/m & 0.5577 & 45.4998 & 0.0000 & 0.0000 & 0.0000 & 0.0625 & 5908.5642 \\
			IC/m & 9.5366 & 128.2722 & 0.0000 & 0.4105 & 1.2044 & 3.7984 & 14601.0533 \\
			PRC/m & 3.3770 & 73.8488 & 0.0000 & 0.0000 & 0.0000 & 0.2847 & 10710.8213 \\
			PRAR & 0.9128 & 0.1588 & 0.0000 & 0.8889 & 1.0000 & 1.0000 & 1.0000 \\
			IRR & 0.6628 & 0.2943 & 0.0000 & 0.5000 & 0.7143 & 0.9200 & 1.0000 \\
			FK/m & 3.0561 & 82.5627 & 0.0000 & 0.1696 & 0.3753 & 0.9217 & 8462.3200 \\
			STR/m & 10.1078 & 147.8087 & 0.0680 & 0.4844 & 1.0854 & 3.1103 & 19207.6400 \\
			\bottomrule
		\end{tabular}
	}
\end{table}

Notably, the maximum values of several metrics were inflated due to the normalization process. For example, commits per month (CPM) reaches 11{,}794{,}250.0000, exceeding the raw commit count maximum of 692{,}245. This is not a coding error but a mathematical consequence of normalization. Specifically, repositories with extremely short lifespans (e.g., 10–15 days) and high commit activity yield inflated per-month rates. The normalization formula applied uniformly across metrics is:

\[
\text{CPM} = \frac{\text{commits}}{\text{active\_lifespan\_days} / 30.44}
\]

For instance, a repository with 700 commits over 15 days yields a CPM of approximately 1{,}419, reflecting burst-like activity rather than sustained development. These outliers were retained to capture the full behavioral spectrum and avoid survivor bias from excluding short-lived projects.

Motivated by these patterns, we visually compared empirical with theoretical distributions using quantile-quantile (QQ) plots (Fig.S1, supplementary material) and analyzed the cumulative distribution function (CDF) of each variable (Fig.S2, supplementary material). While some variables, like \textit{RR} and \textit{PRAR}, appeared less skewed in QQ plots, visual inspection of CDF plots (Fig.S2-S8, supplementary material) generally suggested deviations from a strict power-law distribution.

The power-law distribution, $ f\left(x\right)=ax^{-b} $, models heavy-tailed data where $x$ is a variable, and $a$ and $b$ are constants. Lower $b$ values lead to a slower CDF decline and longer tails, indicating a higher probability of larger values.

\begin{table*}[!h]
	\centering
	\caption{Kolmogorov–Smirnov Test Statistics and Fitted Parameters for Log-Normal and Exponential Distributions}
	\label{tab:ks_test_results}
	\resizebox{\linewidth}{!}{
		\begin{tabular}{l | c c | c c}
			\toprule
			\multirow{2}{*}{\textbf{Attribute}} & \multicolumn{2}{c|}{\textbf{Log-Normal Fit}} & \multicolumn{2}{c}{\textbf{Exponential Fit}} \\
			& KS Statistic & Parameters & KS Statistic & Parameters \\
			\midrule
			CPM & 0.0434 & $s=1.59,\ loc=0.10,\ scale=4.01$ & 0.8451 & $loc=0.10,\ scale=507.88$ \\
			WT/m & 0.0324 & $s=1.39,\ loc=0.01,\ scale=0.23$ & 0.3799 & $loc=0.00,\ scale=1.10$ \\
			CNT/m & 0.0592 & $s=1.29,\ loc=0.02,\ scale=0.18$ & 0.6703 & $loc=0.02,\ scale=2.77$ \\
			TI/m & 0.0198 & $s=1.67,\ loc=0.01,\ scale=0.38$ & 0.3554 & $loc=0.01,\ scale=1.91$ \\
			OI/m & 0.0198 & $s=1.62,\ loc=0.01,\ scale=0.15$ & 0.4042 & $loc=0.00,\ scale=0.66$ \\
			TPR/m & 0.0409 & $s=1.64,\ loc=0.01,\ scale=0.48$ & 0.4156 & $loc=0.01,\ scale=2.89$ \\
			OPR/m & 0.0329 & $s=1.50,\ loc=0.01,\ scale=0.07$ & 0.6529 & $loc=0.00,\ scale=0.56$ \\
			IC/m & 0.0255 & $s=1.73,\ loc=0.01,\ scale=1.34$ & 0.4405 & $loc=0.00,\ scale=9.54$ \\
			PRC/m & 0.0420 & $s=2.14,\ loc=0.01,\ scale=0.41$ & 0.6715 & $loc=0.00,\ scale=3.38$ \\
			FK/m & 0.0476 & $s=1.43,\ loc=0.01,\ scale=0.42$ & 0.4908 & $loc=0.00,\ scale=3.06$ \\
			STR/m & 0.0565 & $s=1.59,\ loc=0.07,\ scale=1.25$ & 0.4939 & $loc=0.07,\ scale=10.04$ \\
			PRAR & 0.2826 & $s=0.00,\ loc=-32752.93,\ scale=32753.84$ & 0.4498 & $loc=0.00,\ scale=0.91$ \\
			RR & 0.1165 & $s=0.00,\ loc=-32750.05,\ scale=32750.76$ & 0.2926 & $loc=0.00,\ scale=0.66$ \\
			\bottomrule
		\end{tabular}
	}
\end{table*}

To further evaluate distributional fit, we applied the Kolmogorov–Smirnov (KS) test to each normalized attribute (Table~\ref{tab:ks_test_results}). While some attributes exhibit modest KS statistics under log-normal fitting (e.g., TI/m, IC/m), others (such as \texttt{PRAR} and \texttt{RR}) yield extreme parameter values, suggesting instability in the fitting process. Exponential fits generally result in higher KS values, with less plausible scale parameters. Though not shown in this table, additional tests with Pareto and Burr distributions produced shape parameters (e.g., $b<2$), indicative of heavy-tailed behavior, yet failed to achieve sufficient goodness-of-fit. These findings reinforce that real-world OSS data often deviates from idealized theoretical models.

\paragraph{Exploratory Factor Analysis}\label{subsubsec:efa}
The EFA process began with preliminary assessments to ensure the suitability of the data for factor extraction. This involved applying the Kaiser-Meyer-Olkin (KMO) measure of sampling adequacy and Bartlett's test of sphericity.

\begin{table}[!h]
	\centering
	\caption{KMO Results and VIF Scores for Individual Attributes}
	\label{tab:kmo_vif_table}
	\setlength{\tabcolsep}{5pt}
	\begin{tabular}{l r r}
		\toprule
		\textbf{Attribute} & \textbf{MSA (KMO)} & \textbf{VIF} \\
		\midrule
		CPM   & \textcolor{gray}{0.48} & 42.0859 \\
		WT/m   & 0.80 & 1.7799 \\
		CNT/m  & \textcolor{gray}{0.49} & 42.7682 \\
		TI/m   & 0.59 & 7.2972 \\
		OI/m   & 0.58 & 7.5078 \\
		TPR/m  & 0.55 & 19.1740 \\
		OPR/m  & 0.53 & 19.8470 \\
		IC/m   & 0.56 & 2.5996 \\
		PRC/m  & \textcolor{gray}{0.48} & 2.2907 \\
		PRAR   & 0.56 & 1.0792 \\
		RR     & \textcolor{gray}{0.42} & 1.0883 \\
		FK/m   & 0.78 & 2.3138 \\
		STR/m  & 0.66 & 1.5513 \\
		\midrule
		\textbf{Overall MSA} & \textbf{0.57} & \\
		\bottomrule
	\end{tabular}
\end{table}

As shown in Table~\ref{tab:kmo_vif_table}, an initial overall KMO value of 0.57 was observed. Examination of individual KMO values, also known as Measures of Sampling Adequacy (MSA), revealed that several attributes fell below the generally accepted threshold of 0.5 for inclusion in EFA. Specifically, \texttt{CPM}, \texttt{CNT/m}, \texttt{PRC/m}, and \texttt{RR} exhibited inadequate MSA values (highlighted in gray in Table~\ref{tab:kmo_vif_table}) and were systematically removed from the dataset.

Following the exclusion of these four attributes, the KMO measure was re-evaluated. The overall KMO value subsequently increased to 0.58, and all remaining variables displayed individual MSA values above the 0.5 threshold, confirming adequate sampling. This refined variable set was then subjected to Bartlett’s test of sphericity, which yielded a highly significant Chi-square value of 372{,}759 ($df = 78$, $p < 0.05$), confirming that the retained variables were sufficiently correlated to warrant factor analysis.

Furthermore, a Variance Inflation Factor (VIF) analysis was performed to identify and address multicollinearity among the attributes. VIF values, also included in Table~\ref{tab:kmo_vif_table}, help detect highly correlated variables that can inflate standard errors and distort factor interpretations. Attributes with VIF scores exceeding typical thresholds (e.g., 5 or 10) were flagged for removal. Specifically, \texttt{CNT/m}, \texttt{CPM}, \texttt{OPR/m}, and \texttt{TPR/m} exhibited very high VIF values, indicating severe multicollinearity and were excluded to ensure a robust factor structure. Notably, \texttt{CPM} and \texttt{CNT/m} had already been removed based on KMO results, reinforcing their exclusion.

After these sequential filtering steps based on KMO and VIF criteria, the dataset was refined to retain four attributes for factor extraction: \texttt{TI/m}, \texttt{IC/m}, \texttt{WT/m}, and \texttt{STR/m}. This final set achieved an overall KMO value of 0.58, with all individual attributes exceeding the 0.5 threshold, confirming the dataset's suitability for EFA.

\begin{figure}[!h]
	\centering
	\includegraphics[width=\linewidth]{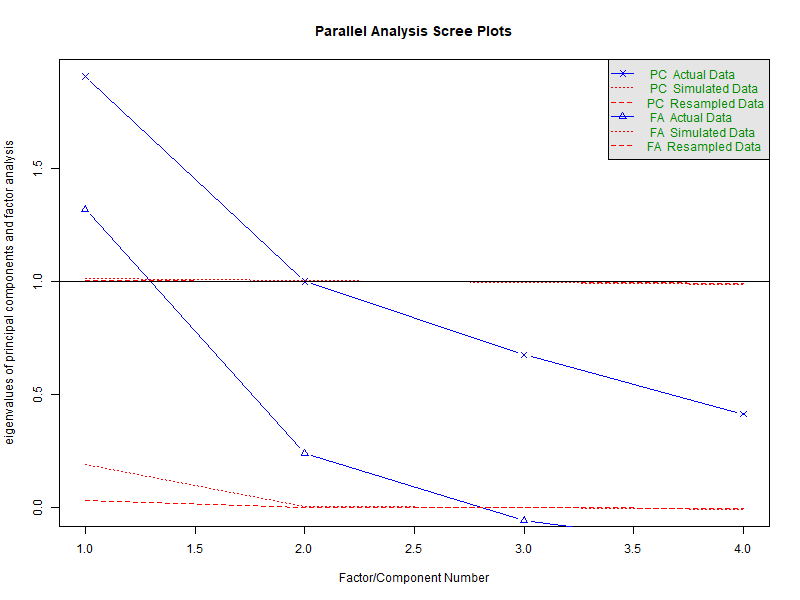}
	\caption{Parallel Analysis Plot}
	\label{fig:parallellanalysisplot}
\end{figure}

Parallel analysis, a robust method, was used to determine the number of factors, suggesting two factors, consistent with both eigenvalues and the scree plot (Figure~\ref{fig:parallellanalysisplot}).

The finalized two-factor solution for the four key engagement indicators is presented comprehensively in Table~\ref{tab:tab16}, including unrotated and rotated factor loadings, communalities ( $h^2$ ), uniquenesses ( $u^2$ ), Hoffman's Index of Complexity (\textit{com}), factor statistics, and model fit. The unrotated loadings, derived using the minimum residual (minres) method, reveal initial correlations between attributes and the two latent factors. To enhance interpretability, Varimax rotation was applied, resulting in clearer loading patterns while preserving communality values.
\begin{table}[!h]
	\centering
	\caption{Factor Loadings, Complexity, and Fit Statistics of Refined Model}
	\label{tab:tab16}
	\resizebox{\linewidth}{!}{
		\begin{tabular}{l c c c c c}
			\toprule
			\textbf{Attribute} & \textbf{MR1} & \textbf{MR2} & \textbf{$h^2$} & \textbf{$u^2$} & \textbf{com} \\
			\midrule
			\multicolumn{6}{c}{\textbf{Unrotated Loadings}} \\
			\midrule
			TI/m  & 0.64 & 0.30 & 0.50 & 0.50 & 1.4 \\
			IC/m  & 0.32 & 0.52 & 0.37 & 0.63 & 1.7 \\
			WT/m  & 0.93 & -0.35 & 0.99 & 0.01 & 1.3 \\
			STR/m & 0.42 & -0.07 & 0.18 & 0.82 & 1.1 \\
			\midrule
			\multicolumn{6}{c}{\textbf{Rotated Loadings (Varimax)}} \\
			\midrule
			WT/m  & 0.99 & 0.11 & 0.99 & 0.01 & 1.0 \\
			TI/m  & 0.44 & 0.56 & 0.50 & 0.50 & 1.9 \\
			IC/m  & 0.05 & 0.61 & 0.37 & 0.63 & 1.0 \\
			STR/m & 0.41 & 0.13 & 0.18 & 0.82 & 1.2 \\
			\midrule
			\multicolumn{6}{c}{\textbf{Factor Statistics (SS Loadings and Variance Explained)}} \\
			\midrule
			\textbf{Factor} & \textbf{SS Loadings} & \textbf{Proportion Var} & \textbf{Cumulative Var} & & \\
			MR1 & 1.35 & 0.34 & 0.34 & & \\
			MR2 & 0.71 & 0.18 & 0.51 & & \\
			\midrule
			\multicolumn{6}{c}{\textbf{Model Fit Statistics}} \\
			\midrule
			\multicolumn{6}{l}{TLI: 1.00 \quad RMSEA: 0.00 \quad SRMR: 0.00} \\
			\bottomrule
		\end{tabular}
	}
\end{table}

The finalized solution demonstrates strong fit indices (TLI = 1.00, RMSEA = 0.00, SRMR = 0.00), confirming excellent factorial reliability and minimal residual error. Communalities remained robust, and most variables exhibited distinct primary loadings exceeding the 0.5 threshold, with the exception of \texttt{STR/m} (0.41), which was retained due to its theoretical relevance and conceptual grouping within its factor. These results confirm the suitability of the retained attributes for modeling community engagement.

\begin{figure}[!h]
	\centering
	\includegraphics[width=\linewidth]{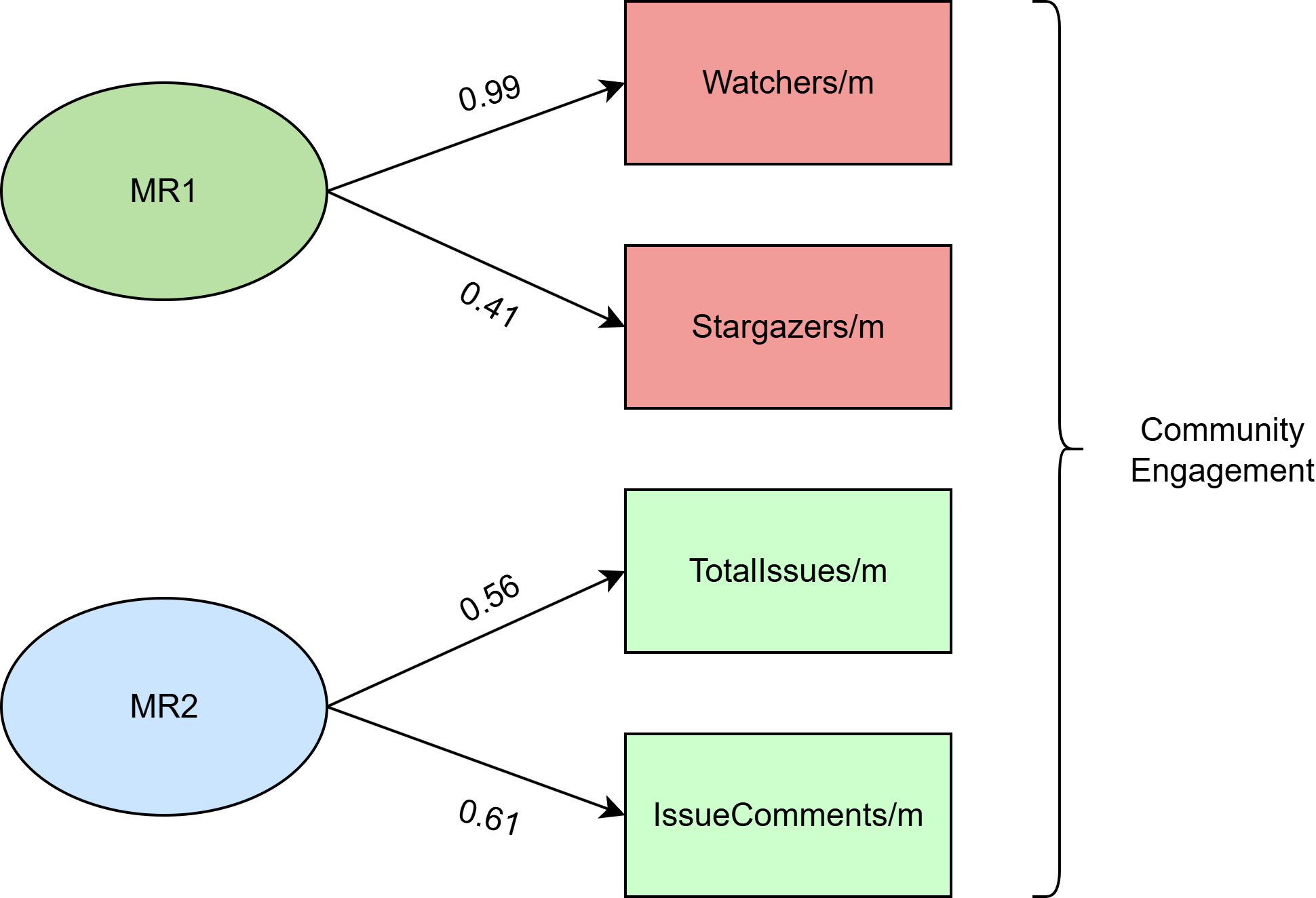}
	\caption{Factor Analysis Diagram: CE Model}
	\label{fig:efamodel}
\end{figure}

Based on the refined EFA model, MR1 groups Watchers per month (WT/m) and Stargazers per month (STR/m), while MR2 is primarily composed of Total Issues per month (TI/m) and Issue Comments per month (IC/m). Drawing on prior research~\cite{Nijsse2023, Tamburri2019}, MR1 is interpreted as \textit{Passive Engagement (PE)}, reflecting external interest and visibility. MR2 represents \textit{Active Engagement (AE)}, capturing direct participation and feedback within OSS communities . The visual structure in Figure~\ref{fig:efamodel} illustrates these dimensions, where TI and IC signify active communication and issue tracking, while WT and STR denote public attention and potential future involvement~\cite{Nijsse2023, Tamburri2019}. Although passive metrics such as WT and STR may precede contribution~\cite{Sheoran2014}, they primarily reflect observational engagement. Issues and comments continue to be strong indicators of collaborative activity and contributor interaction~\cite{bertram2010communication}.

Following the identification and interpretation of the two latent factors, factor scores were computed for each project using the regression method. These scores quantify each project's standing on the derived engagement dimensions. Specifically, the factor scores correspond to Passive Engagement (derived from Watchers and Stargazers) and Active Engagement (derived from Total Issues and Issue Comments). These individual factor scores serve as the primary quantitative measures of community engagement in subsequent analyses of project dynamics and lifespan.

While our primary focus is on the direct, interactive aspects of CE that drive collaborative development (captured by the Active Engagement factor), the exploratory factor analysis also identified a distinct Passive Engagement factor (comprising Watchers per month and Stargazers per month). To provide a comprehensive understanding of how different facets of CE relate to various project attributes, both Active and Passive Engagement factors will be utilized in the subsequent analyses concerning project dynamics (RQ2) and project lifespan (RQ3).

\subsubsection{Cross-validation}\label{subsubsec:cv}

To assess model generalizability, we performed cross-validation by randomly splitting the dataset into a 70\% training subset and a 30\% testing subset. Table~\ref{tab:tab18} summarizes the standardized loadings and key fit statistics across these splits. For interpretive clarity, loadings below 0.5 are not shown.

\begin{table}[!h]
	\centering
	\caption{Cross-validation Results}
	\label{tab:tab18}
	\resizebox{\linewidth}{!}{
		\begin{tabular}{l c c c c c c}
			\toprule
			& \multicolumn{2}{c}{\textbf{Overall Fit}} & \multicolumn{2}{c}{\textbf{Training}} & \multicolumn{2}{c}{\textbf{Testing}} \\
			\midrule
			\textbf{Attribute} & \textbf{AE} & \textbf{PE} & \textbf{AE} & \textbf{PE} & \textbf{AE} & \textbf{PE} \\
			TI/m   & 0.56 & ---  & 1.00 & ---  & 1.00 & ---  \\
			IC/m   & 0.61 & ---  & 0.35 & ---  & 0.49 & ---  \\
			WT/m   & ---  & 0.99 & ---  & 0.71 & ---  & 0.91 \\
			STR/m  & ---  & 0.41 & ---  & 0.38 & ---  & 0.80 \\
			\midrule
			\textbf{n}     & \multicolumn{2}{c}{33,946} & \multicolumn{2}{c}{23,764} & \multicolumn{2}{c}{10,182} \\
			\textbf{TLI}   & \multicolumn{2}{c}{1.00}   & \multicolumn{2}{c}{0.953}  & \multicolumn{2}{c}{0.993}  \\
			\textbf{RMSEA} & \multicolumn{2}{c}{0.00}   & \multicolumn{2}{c}{0.067}  & \multicolumn{2}{c}{0.036}  \\
			\textbf{SRMR}  & \multicolumn{2}{c}{0.012}  & \multicolumn{2}{c}{0.020}  & \multicolumn{2}{c}{0.012}  \\
			\textbf{CFI}   & \multicolumn{2}{c}{1.00}   & \multicolumn{2}{c}{0.984}  & \multicolumn{2}{c}{0.998}  \\
			\bottomrule
		\end{tabular}
	}
\end{table}

Results confirm robust factorial stability across data subsets. Loadings for Total Issues (TI/m) and Watchers (WT/m) remained strong and consistent, reinforcing the latent dimensions of Active and Passive Engagement respectively. Goodness-of-fit indices (e.g., TLI and CFI approaching 1, RMSEA below 0.08~\cite{Xia2019}) support model adequacy in both training and testing contexts. Minor variability in loadings and fit indices is expected due to sample fluctuations, but does not compromise model integrity.

\begin{minipage}{0.95\linewidth}
	\hrule
	\vspace{5pt}
	\noindent \textbf{Answer to RQ.1(b):} Through exploratory factor modeling and cross-validation, we empirically confirmed that observable development and attention-related metrics converge into two latent dimensions of Community Engagement. Specifically, Total Issues and Issue Comments align with \textit{Active Engagement}, denoting direct participation and collaborative effort. Meanwhile, Watchers and Stargazers correspond to \textit{Passive Engagement}, signaling public interest and project visibility. This bifurcation provides a validated structural framework for conceptualizing CE in OSS ecosystems.
	\vspace{5pt}
	\hrule
\end{minipage}

\subsection{Metrics Validation}\label{Metrics Validation}
\subsubsection{Mapping EFA Factors to CE Definition}\label{MapEFACE}

Building on our comprehensive CE definition in OSS (Section~\ref{subsec:DCE}) and the empirical identification of its underlying dimensions through EFA, this section validates the construct validity of our CE model. EFA revealed two primary factors: \textit{Passive Engagement} (MR1) and \textit{Active Engagement} (MR2). The following mapping illustrates how these empirically derived factors and their constituent attributes conceptually align with our broader CE definition, ensuring that the statistical model accurately reflects the intended theoretical construct.

\begin{itemize}
	\item \textbf{Passive Engagement (MR1)} consists of \texttt{Watchers per month} and \texttt{Stargazers per month}, which indicate a broader, less direct level of community interest. Although these individuals may not actively contribute through discussions or code, their presence reflects an audience observing the project and signals potential future contributors~\cite{Sheoran2014}. A substantial number of watchers and stargazers enhances project visibility and supports the notion of a sustainable community by serving as a leading indicator of prospective engagement.
	
	\item \textbf{Active Engagement (MR2)} includes \texttt{Total Issues per month} and \texttt{Issue Comments per month}, directly capturing dynamic participation through structured communication. A high volume of issues suggests frequent use of the issue tracker by diverse stakeholders, particularly users, for bug reporting and feedback, emphasizing direct forms of collaboration~\cite{bertram2010communication}. Rich comment threads further reflect sustained use of communication channels for resolving problems collectively, strengthening collaborative relationships that contribute to OSS quality and community cohesion~\cite{rath2020request}.
\end{itemize}

\subsubsection{Measurement Scales and Metric Interpretation}\label{TVCEM}

This section evaluates the proposed CE metrics: \texttt{Total Issues per month (TI/m)}, \texttt{Issue Comments per month (IC/m)}, \texttt{Watchers per month (WT/m)}, and \texttt{Stargazers per month (STR/m)}. We examine their validity and suitability in capturing various aspects of CE, beginning with their underlying measurement principles.

Measurement involves assigning numbers to objects or events based on defined rules, with interpretation varying by scale. Our metrics primarily operate on a ratio scale, meaning they possess a true zero point, allowing for meaningful ratio comparisons (e.g., twice the issues per month indicates double the issue-related activity relative to lifespan). It's crucial to note that all selected metrics were normalized by the project's active lifespan (i.e., calculated 'per month') to account for project longevity and enable fair comparisons across projects of varying durations. However, it's important to distinguish this quantitative aspect from the qualitative interpretation of CE; for instance, twice the issues per month doesn't necessarily imply twice the quality of engagement.

Metric interpretation is highly context-dependent. A single metric can reflect various underlying attributes depending on its application. For example, \texttt{Total Issues per month} can represent development activity (e.g., bug reports) or, in our context, CE by indicating active use of the issue tracker. This highlights the critical need to consider the specific research question when interpreting our metrics.

Specifically, these metrics quantify aspects of Active and Passive Community Engagement:

\begin{itemize}
	\item \textbf{Active Engagement Metrics:}
	\begin{itemize}
		\item \textbf{Total Issues per month (TI/m) and Issue Comments per month (IC/m):} These metrics offer insight into the level of active participation within the project's issue tracking system, reflecting the extent of CE in discussions, bug reports, and feature proposals. \texttt{Issue Comments per month} specifically captures the degree of interaction and feedback within issue discussions, measuring collaborative problem-solving and knowledge sharing.
	\end{itemize}
	\item \textbf{Passive Engagement Metrics:}
	\begin{itemize}
		\item \textbf{Watchers per month (WT/m) and Stargazers per month (STR/m):} These metrics reflect broader community interest and project visibility. They indicate the size of the audience following the project, serving as indicators of potential future engagement and contributing to the project's sustainable community.
	\end{itemize}
\end{itemize}

It's important to acknowledge that CE is a multifaceted concept. Our metrics capture specific aspects of observable activity and attention but don't fully represent underlying motivations, influence, or qualitative aspects of interactions. This limitation offers avenues for future research.

\subsubsection{Formal Validation using Weyuker's Properties}

After clarifying our metrics' measurement scales and interpretative scope, we formally evaluate their validity and limitations against Weyuker's properties, adopting the interpretations by Srinivasan et al.~\cite{srinivasan2014software}. We assessed our proposed CE metrics (namely, \texttt{Total Issues per month (TI/m)}, \texttt{Issue Comments per month (IC/m)}, \texttt{Watchers per month (WT/m)}, and \texttt{Stargazers per month (STR/m)}) against these nine properties:

\begin{itemize}
	\item \textbf{Non-coarseness (P1)}: For any project $P$, there exists another project $Q$ such that $\mu(P) \ne \mu(Q)$. Our metrics satisfy this property; it is highly probable to find projects with different metric values, such as the number of issues per month or watchers per month.
	
	\item \textbf{Granularity (P2)}: For any given value $c$, the set of projects $P$ such that $\mu(P) = c$ is finite. This property is satisfied, as the number of projects and their corresponding metric values (e.g., issues per month, comments per month, watchers per month, stargazers per month) are finite in practice.
	
	\item \textbf{Nonuniqueness (P3)}: There exist distinct projects $P$ and $Q$ such that $\mu(P) = \mu(Q)$. This property is satisfied, as two different projects can have the same value for a given metric, such as identical numbers of total issues per month.
	
	\item \textbf{Design Details are Important (P4)}: This property is not satisfied. Our metrics are insensitive to how the community engages (such as interaction quality, issue context, or the nature of a watch or star), and therefore may assign similar values to projects with vastly different engagement profiles.

	\item \textbf{Monotonicity (P5)}: If $P+Q$ denotes combining projects, then $\mu(P+Q) \ge \mu(P)$ and $\mu(P+Q) \ge \mu(Q)$. This condition is generally met; adding engagement activity (e.g., issues, comments) or public interest (e.g., watchers, stargazers) tends to increase aggregate metric values.
	
	\item \textbf{Nonequivalence of Interaction (P6)}: Even if $\mu(P) = \mu(Q)$, it is not guaranteed that $\mu(P+R) = \mu(Q+R)$. This property is not satisfied; our metrics do not account for how an additional project or contextual factor $R$ may interact differently with $P$ and $Q$ to influence engagement.
	
	\item \textbf{Permutation (P7)}: If $S'$ is a permutation of the event sequence $S$, then $\mu(S) = \mu(S')$. This property is not satisfied, as our metrics are order-insensitive (e.g., Issue Comments per month remain the same regardless of when comments are posted).
	
	\item \textbf{Renaming Property (P8)}: If project $P'$ is a renamed version of project $P$, then $\mu(P') = \mu(P)$. This property is satisfied by our count-based metrics, which remain invariant under project renaming since the underlying activity and attention are unchanged.
	
	\item \textbf{Interaction Increases Complexity (P9)}: $\mu(P+Q) > \mu(P) + \mu(Q)$. This property is not strictly satisfied. While aggregate engagement may numerically exceed the sum of individual metrics due to synergistic effects, our metrics do not explicitly model the added qualitative complexity from community interactions.
\end{itemize}

Our proposed metrics fail to satisfy properties P4, P6, and P7, directly limiting our CE interpretation. These limitations stem from their focus on activity volume and attention over interaction quality, context, or dynamics. Specifically, they are insensitive to design details (P4), fail to capture interaction nuances (P6), and do not consider event order (P7). For example, a project with numerous superficial comments per month might appear more actively engaged than one with fewer, more insightful discussions. This lack of sensitivity prevents understanding how different interactions qualitatively influence project health or community evolution.

Nevertheless, failing all of Weyuker's properties doesn't invalidate metrics. Their applicability varies by context and analytical goals~\cite{misra2008applicability}. Empirical evidence shows metrics not fully adhering to all properties, like McCabe's cyclomatic complexity, can still yield useful information~\cite{de2013applicability}. Our choice to prioritize readily available, automatically collectible data led us to focus on metrics capturing activity volume and visible interest. While this sacrifices sensitivity to interaction dynamics and quality, it provides a valuable baseline understanding of CE and enables robust longitudinal analysis. Despite these acknowledged limitations, our metrics offer insights into interaction and contribution volume, forming a solid foundation for comparing projects or tracking changes over time. The significance of each Weyuker property ultimately depends on specific research objectives; for instance, capturing activity volume (P5) might be more critical than sensitivity to event order (P7) when tracking community growth.

\subsubsection{Face Validity Check}\label{face_validity}

We conducted a face validity check to complement the formal validation by Weyuker's properties, assessing the proposed CE metrics' practical relevance and interpretability. This aimed to gather expert opinions on whether the defined metrics effectively capture the intended CE aspects.

The check utilized a questionnaire (see Supplementary Material, Section 2) distributed to two groups: (1) 10 OSS users and researchers, and (2) 6 OSS maintainers and experts. Participants were selected based on extensive experience in OSS development, community management, or software engineering research, ensuring diverse perspectives. The questionnaire presented our CE definition, metric-to-aspect mapping, and included general relevance questions, specific 5-point Likert scale ratings for relevance and interpretability, and optional comment fields. A pilot test with five participants confirmed the questionnaire's clarity, requiring no significant revisions.

\begin{table}[!h]
	\centering
	\caption{Mean and Standard Deviation of Relevance and Interpretability Ratings for CE Metrics}
	\label{tab:face_validity_combined}
	\resizebox{\linewidth}{!}{
		\begin{tabular}{l l c c c c}
			\toprule
			\textbf{Metric} & \textbf{Rating Type} & \multicolumn{2}{c}{\textbf{Group 1}} & \multicolumn{2}{c}{\textbf{Group 2}} \\
			\cmidrule(lr){3-4} \cmidrule(lr){5-6}
			& & Mean & Std Dev & Mean & Std Dev \\
			\midrule
			\multirow{2}{*}{TotalIssues/m} 
			& Relevance         & 4.6 & 0.52 & 4.5 & 0.55 \\
			& Interpretability  & 4.5 & 0.67 & 4.5 & 0.55 \\
			
			\multirow{2}{*}{IssueComments/m}
			& Relevance         & 4.7 & 0.48 & 4.7 & 0.48 \\
			& Interpretability  & 4.7 & 0.48 & 4.8 & 0.41 \\
			
			\multirow{2}{*}{Watchers/m}
			& Relevance         & 4.5 & 0.53 & 4.5 & 0.55 \\
			& Interpretability  & 4.3 & 0.48 & 4.3 & 0.41 \\
			
			\multirow{2}{*}{Stargazers/m}
			& Relevance         & 4.5 & 0.53 & 4.5 & 0.55 \\
			& Interpretability  & 4.3 & 0.48 & 4.3 & 0.41 \\
			\bottomrule
		\end{tabular}
	}
\end{table}

Table \ref{tab:face_validity_combined} presents descriptive statistics for relevance and interpretability ratings from both groups for all four normalized metrics. All metrics consistently received high ratings. For Total Issues/m, Issue Comments/m, Watchers/m, and Stargazers/m, mean relevance scores ranged from 4.5 to 4.7 for both groups, and mean interpretability scores ranged from 4.3 to 4.7 for Group 1 and 4.3 to 4.8 for Group 2. Notably, Issue Comments/m received consistently high ratings across both groups (mean relevance 4.7; mean interpretability Group 1: 4.7, Group 2: 4.8), indicating strong agreement on its suitability and clarity. Despite minor variations, the consistent high ratings strongly support the relevance and interpretability of the selected CE metrics, providing confidence in their validity for subsequent quantitative analysis.

\section{Community Engagement and Project Dynamics}\label{sec:CEPD}

This section addresses RQ.~\ref{RQ2} by investigating CE's influence on project dynamics. We define `project dynamics' through three key indicators: commits, branches, and releases. These metrics collectively reflect OSS project evolutionary stages, from planning to stable releases~\cite{Saraf2013}. Releases mark significant milestones, achieved through continuous developer contributions via commits, while branches represent individual development pathways. We begin by exploring correlations between CE factors and these dynamics, setting the stage for a focused analysis on CE's influence on commit activity. For clarity, `CE factors' in this study refer to the composite Active and Passive Engagement Scores empirically derived from Factor Analysis, while `CE metrics' denote the individual underlying data points (e.g., total issues, watchers) that comprise these factors.

\subsection{Correlation between CE Factors and Project Dynamics}\label{CorrCEPD}
\subsubsection{Methods}\label{CEPDM}

Recognizing a gap in prior research, we employed correlation coefficients and other relevant statistical tests, implemented using Python, to investigate the relationship between CE factors and project dynamics. These CE factors, operationalized as Active and Passive Engagement Scores (as identified in Section~\ref{sec:DMCE}), represent empirically validated latent dimensions of community engagement, allowing for a more holistic and robust assessment of engagement's influence on project dynamics than individual metrics alone. Given the varying ages of projects and to ensure consistent comparison with our normalized CE metrics, project dynamics (commits, branches, and releases) were also normalized by the project's active lifespan, yielding metrics such as commits\_per\_month. Due to observed deviations from normality in the data distributions, Spearman's rank correlation was chosen to assess these relationships.

\subsubsection{Results}\label{CEPD2aR}

Before conducting correlation analysis, variable distributions were examined. CE factor distributions were discussed in Section~\ref{RDD}, while descriptive statistics for normalized project dynamics (namely, \texttt{commits\_per\_month (CPM)}, \texttt{branches\_per\_month (BPM)}, and \texttt{releases\_per\_month (RPM)}) are reported in Table~\ref{tab:tab21}.

\begin{table}[!h]
	\centering
	\caption{Descriptive Statistics of Normalized Project Dynamics}
	\label{tab:tab21}
		\begin{tabular}{l c c r}
			\toprule
			& \textbf{CPM} & \textbf{BPM} & \textbf{RPM} \\
			\midrule
			Mean        & 507.98         & 0.66           & 0.53           \\
			Std Dev     & 64996.92       & 19.08          & 10.32          \\
			Skewness    & 176.624        & 132.025        & 101.236        \\
			Kurtosis    & 31950.506      & 19531.611      & 11588.962      \\
			Min         & 0.10           & 0.01           & 0.01           \\
			25\%        & 1.44           & 0.04           & 0.03           \\
			Median      & 3.53           & 0.08           & 0.07           \\
			75\%        & 10.41          & 0.23           & 0.26           \\
			Max         & $1.18 \times 10^{7}$ & 3013.56        & 1369.80        \\
			\bottomrule
		\end{tabular}
\end{table}

The distribution patterns in Table~\ref{tab:tab21} exhibit considerable deviation from normality, reaffirming the choice of Spearman’s rank correlation for robustness. The resulting analysis demonstrates strong and statistically significant associations between CE scores and project dynamics. Specifically:
\begin{itemize}
	\item \textbf{AES} shows positive correlations with:
	\begin{itemize}
		\item \texttt{CPM}: $\rho = 0.664$
		\item \texttt{BPM}: $\rho = 0.548$
		\item \texttt{RPM}: $\rho = 0.448$
	\end{itemize}
	
	\item \textbf{PES} is also positively correlated with:
	\begin{itemize}
		\item \texttt{CPM}: $\rho = 0.495$
		\item \texttt{BPM}: $\rho = 0.496$
		\item \texttt{RPM}: $\rho = 0.359$
	\end{itemize}
\end{itemize}

All reported correlations are significant at $p < 0.001$, underscored by observed $p$-values $\approx 0.0000$. These relationships indicate that increased levels of both active and passive community engagement are closely associated with elevated repository dynamics across diverse technical activities. Nonetheless, as correlation analysis reflects statistical associations rather than causal direction, further investigation into potential underlying mechanisms or moderating factors is warranted.

\begin{minipage}[t]{0.95\linewidth}
	\hrule
	\vspace{5pt}
	\noindent \textbf{Answer to RQ.2(a):}
	
	\noindent Both \textit{AES} and \textit{PES} exhibit strong, statistically significant correlations with project dynamics (specifically, commits, branches, and RPM), confirming their relevance as indicators of OSS activity levels.
	
	\vspace{5pt}
	\hrule
\end{minipage}

\subsection{Varying Correlation in High or Low Active Engagement Projects}\label{subsubsec:VaryCorr}

\subsubsection{Methods}\label{RQ2Methods}

To examine how project engagement levels influence the relationship between CE factors and project dynamics, we employed a multi-stage approach. First, we divided the dataset at the median AES to classify projects as either low or high active engagement, given the absence of established thresholds. We focused on AES because its metrics (issues, comments) directly reflect internal, development-driving activity and show stronger correlations with project dynamics than passive engagement. Next, we computed Spearman's rank correlation coefficients ($\rho$) separately for both groups. To evaluate the statistical significance of differences in these coefficients between engagement levels, we will use bootstrap resampling (Algorithm~\ref{alg:bootstrap_spearman}), generating multiple samples by randomly selecting data points with replacement to enhance reliability. Z-scores and p-values will then be calculated to verify coefficient differences.

\begin{algorithm}
	\caption{Bootstrapping Procedure to Compare Spearman's Rank Correlation Coefficients}
	\label{alg:bootstrap_spearman}
	\begin{algorithmic}[1]
		\Require $X$, $Y$ \Comment{Data for the `high active engagement' and `low active engagement' groups, respectively}
		\Ensure $z$, $p$ \Comment{z-score and p-value for the difference in Spearman's rho}
		\State $\rho_{X,Y} \gets \text{Spearmanr}(X, Y)$ \Comment{Calculate the original Spearman's rho}
		\State $B \gets 10,000$ \Comment{Number of bootstrap iterations}
		\State $\Delta\rho \gets []$ \Comment{List to store the bootstrap differences}
		\For{$i \gets 1$ to $B$}
		\State $X_i \gets \text{sample}(X, |X|, \text{replace=True})$ \Comment{Draw a bootstrap sample from $X$}
		\State $Y_i \gets \text{sample}(Y, |Y|, \text{replace=True})$ \Comment{Draw a bootstrap sample from $Y$}
		\State $\rho_i \gets \text{Spearmanr}(X_i, Y_i)$ \Comment{Calculate Spearman's rho for the bootstrap samples}
		\State $\Delta\rho_i \gets \rho_i - \rho_{X,Y}$ \Comment{Calculate the difference from the original Spearman's rho}
		\State Append $\Delta\rho_i$ to $\Delta\rho$ \hspace{5pt} \Comment{Add the difference in the list}    
		\EndFor
		\State $\text{SE}_\text{bootstrap} \gets \text{std}(\Delta\rho)$ \Comment{Calculate the bootstrap standard error}
		\State $z \gets \frac{\rho_{X,Y}}{\text{SE}_\text{bootstrap}}$ \Comment{Calculate the z-score}
		\State $p \gets 2 \cdot (1 - \Phi(|z|))$ \Comment{Calculate the p-value}	
		\Return $z$, $p$
	\end{algorithmic}
\end{algorithm}

\subsubsection{Results}

For this analysis, we chose the AES to classify project engagement levels. This factor, derived from direct interactive contributions, provides a robust and continuous measure of a project's core engagement. The dataset was divided at the 50th percentile of the AES -0.0906, classifying projects below this as low active engagement and those above as high active engagement. Descriptive statistics for the overall project dynamics are presented in Table~\ref{tab:tab21}.

\begin{table}[!h]
	\centering
	\caption{Spearman Correlation ($\rho$) between Engagement Scores and Project Dynamics}
	\label{tab:tab23}
		\begin{tabular}{l l c c}
			\toprule
			\textbf{Score Type} & \textbf{Metric} & \textbf{Low} & \textbf{High} \\
			\midrule
			\multirow{3}{*}{Active Engagement} 
			& CPM  & 0.344 & 0.578 \\
			& BPM & 0.324 & 0.408 \\
			& RPM & 0.272 & 0.309 \\
			\midrule
			\multirow{3}{*}{Passive Engagement} 
			& CPM  & 0.275 & 0.355 \\
			& BPM & 0.340 & 0.389 \\
			& RPM & 0.196 & 0.261 \\
			\bottomrule
		\end{tabular}
\end{table}

Table~\ref{tab:tab23} reports Spearman correlation coefficients ($\rho$) for both low and high active engagement project groups. Notable differences emerge across key project dynamics:

\begin{itemize}
	\item For \textbf{AES}, correlations with \texttt{commits\_per\_month}, \texttt{branches\_per\_month}, and \texttt{releases\_per\_month} are markedly stronger in the high engagement group (e.g., $\rho = 0.578$ vs $\rho = 0.344$ for commits), suggesting that projects with greater active participation demonstrate tighter alignment with technical development indicators.
	
	\item For \textbf{PES}, correlations also trend higher in the high engagement group across all metrics (most prominently for releases, e.g., $\rho = 0.261$ vs $\rho = 0.196$), indicating that broader project visibility may complement, rather than replace, active contributions in well-engaged projects.
	
\end{itemize}

All correlation coefficients are significant at $p < 0.001$. To formally assess the statistical significance of these observed differences in correlation coefficients between active engagement levels, we used bootstrap resampling (Algorithm~\ref{alg:bootstrap_spearman}), generating 10,000 random samples. The results of this analysis are presented in Table~\ref{tab:tab25}.

\begin{table}[!h]
	\centering
	\caption{Bootstrap Analysis: Correlation Coefficient Differences, Z-scores, and P-values between High and Low Active Engagement Groups}
	\label{tab:tab25}
	\resizebox{\linewidth}{!}{
		\begin{tabular}{p{38pt}p{32pt}p{34pt}p{34pt}p{32pt}p{34pt}p{25pt}}
			\toprule
			\textbf{Metric} & \multicolumn{3}{c}{\textbf{Mean Difference}} & \multicolumn{3}{c}{\textbf{Median Difference}} \\
			\midrule
			& CPM & BPM & RPM & CPM & BPM & RPM \\
			\midrule
			AES & 0.234 & 0.084 & 0.037 & 0.234 & 0.084 & 0.037 \\
			PES & 0.080 & 0.049 & 0.065 & 0.080 & 0.049 & 0.065 \\
			\midrule
			\multicolumn{7}{c}{\textbf{Z-scores and P-values ($p < 0.05$)}} \\
			\midrule
			& CPM & BPM & RPM & CPM & BPM & RPM \\
			\midrule
			AES & 26.124 & 8.680 & 3.602 & 0.000 & 0.000 & 0.000 \\
			PES & 7.906 & 4.956 & 6.176 & $2.67 \times 10^{-15}$ & $7.21 \times 10^{-7}$ & $6.56 \times 10^{-10}$ \\
			\bottomrule
		\end{tabular}
	}
\end{table}

Table~\ref{tab:tab25} reveals that the mean and median differences for each metric are in close proximity, suggesting consistency and minimal variation across the resampled datasets. Furthermore, the standard deviations for these pairwise correlation differences remained low (all below 0.011), indicating that the observed differences are statistically robust.

All pairwise correlation coefficient differences are statistically significant at p<0.001, as evidenced by the extremely low p-values. This strongly suggests that the influence of both Active and Passive Engagement on project dynamics indeed varies significantly depending on the overall engagement level of the project, with generally stronger correlations observed in projects with higher active engagement.

\begin{minipage}[t]{0.95\linewidth}
	\hrule
	\vspace{5pt}
	\noindent \textbf{Answer to RQ. 2(b):}
	\noindent Yes, the analysis demonstrates distinct and statistically significant variations in correlation coefficients between high and low active engagement projects. Bootstrap resampling confirms these differences are consistently observed across resampled data, reinforcing the robustness of the findings. Specifically, the relationships between both Active and Passive Engagement and project dynamics tend to be stronger in projects with higher overall active engagement.
	\vspace{5pt}
	\hrule
\end{minipage}

\subsection{CE Factors influencing Project Dynamics}\label{subsubsec:CEIPD}
Building on the correlation analysis, this section examines how CE factors influence project evolution, specifically commit activity, branch creation, and release frequency.

\subsubsection{Methods}
To assess this relationship, we employed Ordinary Least Squares (OLS) regression on log-transformed dependent variables. This approach was chosen due to the highly skewed, overdispersed, and Pareto-like nature of the project dynamics metrics (namely, \texttt{commits\_per\_month}, \texttt{branches\_per\_month}, and \texttt{releases\_per\_month}), which rendered traditional count models (e.g., Poisson and Negative Binomial GLMs) numerically unstable or unsuitable. The logarithmic transformation helped normalize these distributions, stabilize variance (addressing heteroscedasticity), and linearize multiplicative effects, making the data amenable to OLS analysis. As all original values were strictly positive, a natural logarithm transformation $\ln(x)$ was applied directly to each dependent variable.

Separate OLS regression models were fit for each project dynamic metric, with AES and PES as independent variables. Regression coefficients, $p$-values, and adjusted $R^2$ values were reported to quantify magnitude, significance, and variance explained by CE factors.

\subsubsection{Results}
Table~\ref{tab:glm_results} presents results from the OLS models using log-transformed dependent variables. Across all three models, both AES and PES show statistically significant positive effects on project dynamics ($p < 0.001$).

\begin{table}[h!]
	\centering
	\caption{OLS Regression Results: Influence of CE Factors on Log-Transformed Project Dynamics}
	\label{tab:glm_results}
	\resizebox{\linewidth}{!}{
		\begin{tabular}{l c c c c c c}
			\toprule
			\textbf{Predictor} & \multicolumn{2}{c}{Log(CPM)} & \multicolumn{2}{c}{Log(BPM)} & \multicolumn{2}{c}{Log(RPM)} \\
			\cmidrule(lr){2-3} \cmidrule(lr){4-5} \cmidrule(lr){6-7}
			& Coef. & $p$-value & Coef. & $p$-value & Coef. & $p$-value \\
			\midrule
			Intercept               & 1.4481  & $< 0.001$ & $-2.3030$ & $< 0.001$ & $-2.4171$ & $< 0.001$ \\
			AES                     & 0.4578  & $< 0.001$ & 0.3551   & $< 0.001$ & 0.3389   & $< 0.001$ \\
			PES                     & 0.0719  & $< 0.001$ & 0.1121   & $< 0.001$ & 0.0996   & $< 0.001$ \\
			\midrule
			Adjusted $R^2$          & \multicolumn{2}{c}{0.066} & \multicolumn{2}{c}{0.058} & \multicolumn{2}{c}{0.043} \\
			$N$                     & \multicolumn{2}{c}{33,946} & \multicolumn{2}{c}{33,946} & \multicolumn{2}{c}{33,946} \\
			\bottomrule
		\end{tabular}
	}
\end{table}

\noindent Coefficients can be interpreted as percentage increases using the transformation $\exp(\beta) - 1$:
\begin{itemize}
	\item \textbf{Commits/month (CPM)}:
	AES: $\exp(0.4578)-1 \approx 0.581$ → 58.1\% increase per unit AES increase\\
	PES: $\exp(0.0719)-1 \approx 0.075$ → 7.5\% increase per unit PES increase
	
	\item \textbf{Branches/month (BPM)}:
	AES: $\exp(0.3551)-1 \approx 0.426$ → 42.6\% increase\\
	PES: $\exp(0.1121)-1 \approx 0.119$ → 11.9\% increase
	
	\item \textbf{Releases/month (RPM)}:
	AES: $\exp(0.3389)-1 \approx 0.404$ → 40.4\% increase\\
	PES: $\exp(0.0996)-1 \approx 0.105$ → 10.5\% increase
\end{itemize}

The adjusted R-squared values for the models are 0.066 for CPM, 0.058 for BPM, and 0.043 for RPM. While these R-squared values may appear modest, it is important to contextualize them within the domain of social and organizational systems. In such complex environments, outcomes like project dynamics are influenced by a multitude of factors, many of which are unmeasured or unquantifiable in any given model. Our models aim to identify significant factors of influence rather than to provide high predictive accuracy for exact future values. This perspective is consistent with findings in similar studies of open-source software (OSS); for instance, Jarczyk et al.~\cite{jarczyk2014github} reported an $R 
^2$of 0.203 in their regression model for GitHub project popularity, yet emphasized the interpretability and significance of individual coefficients in understanding project attributes. The consistent and high statistical significance of all coefficients ($p<0.001$) in our models further confirms that both active and passive engagement have a genuine and measurable influence on project dynamics, demonstrating their practical significance despite not explaining the entirety of the variance.

\begin{minipage}[t]{0.95\linewidth}
	\hrule
	\vspace{5pt}
	\noindent \textbf{Answer to RQ. 2(c):} Both AES and PES significantly impact project dynamics. AES is associated with substantial increases in commits, branches, and releases per month, while PES also shows statistically significant (though smaller) positive effects. AES consistently exerts a stronger influence across all three dimensions.
	
	\vspace{5pt}
	\hrule
\end{minipage}

\subsection{Varying relationship between CE Metrics and Project Dynamics with Project Age}\label{subsubsec:CEPDAge}

This section investigates how project age moderates the relationship between individual community engagement (CE) metrics and project dynamics, addressing Research Question (RQ) \ref{RQ2d}.

\subsubsection{Methods}
Project age was operationalized as the difference between project creation date and last release date, and categorized into four quartiles: G1, G2, G3, and G4. This quartile-based grouping was chosen to capture potential non-linear, phase-specific shifts in the relationship between Community Engagement (CE) and project dynamics, by segmenting the data based on its natural distribution.

\subsubsection{Results}

Table \ref{tab:age_interaction_results} presents the OLS regression results. The models showed moderate explanatory power (Adjusted R-squared around 0.38-0.51). The consistent statistical significance of the coefficients, however, demonstrates the potential predictive utility of the individual CE metrics, as they consistently indicate a measurable and predictable influence on project dynamics.

\begin{table*}[!h]
	\centering
	\caption{OLS Regression Results: Influence of Individual CE Metrics on Log-Transformed Project Dynamics, Moderated by Age Group}
	\label{tab:age_interaction_results}
		\begin{tabular}{l c c c c c c}
			\toprule
			\textbf{Predictor (Reference: G1)} & \multicolumn{2}{c}{\textbf{Log(CPM)}} & \multicolumn{2}{c}{\textbf{Log(BPM)}} & \multicolumn{2}{c}{\textbf{Log(RPM)}} \\
			\cmidrule(lr){2-3} \cmidrule(lr){4-5} \cmidrule(lr){6-7}
			& \textbf{Coef.} & \textbf{P-value} & \textbf{Coef.} & \textbf{P-value} & \textbf{Coef.} & \textbf{P-value} \\
			\midrule
			Intercept & 1.3170 & 0.000 & -2.1945 & 0.000 & -2.3073 & 0.000 \\
			
			\multicolumn{7}{l}{\textbf{Main Effects of CE Metrics (G1):}} \\
			Log(totalIssues/month)     & 0.7140 & 0.000 & 0.3785 & 0.000 & 0.2633 & 0.000 \\
			Log(IssueComments/month)   & 0.1077 & 0.000 & 0.1279 & 0.000 & 0.1051 & 0.000 \\
			Log(watchers/month)        & 0.4175 & 0.000 & 0.5432 & 0.000 & 0.3560 & 0.000 \\
			Log(stargazers/month)      & 0.0320 & 0.022 & 0.0945 & 0.000 & 0.2138 & 0.000 \\
			
			\multicolumn{7}{l}{\textbf{Main Effects of Age Groups (vs. G1):}} \\
			G2                & -0.5492 & 0.000 & -0.4327 & 0.000 & -0.2769 & 0.000 \\
			G3                    & -0.9215 & 0.000 & -0.9225 & 0.000 & -0.8248 & 0.000 \\
			G4                      & -1.2130 & 0.000 & -1.4437 & 0.000 & -1.4789 & 0.000 \\
			
			\multicolumn{7}{l}{\textbf{Interaction Terms:}} \\
			TI × G2                        & 0.0537  & 0.180 & -0.1741 & 0.000 & -0.0301 & 0.509 \\
			TI × G3                        & 0.1645  & 0.001 & -0.2215 & 0.000 & -0.0662 & 0.230 \\
			TI × G4                        & 0.1158  & 0.032 & -0.2374 & 0.000 &  0.0711 & 0.248 \\
			IC × G2                        & 0.2450  & 0.000 &  0.2532 & 0.000 &  0.1824 & 0.000 \\
			IC × G3                        & 0.3406  & 0.000 &  0.3341 & 0.000 &  0.3294 & 0.000 \\
			IC × G4                        & 0.4888  & 0.000 &  0.3984 & 0.000 &  0.3457 & 0.000 \\
			WT × G2                        & -0.1175 & 0.016 & -0.2770 & 0.000 & -0.4250 & 0.000 \\
			WT × G3                        & -0.0566 & 0.327 & -0.0997 & 0.061 & -0.4873 & 0.000 \\
			WT × G4                        &  0.1272 & 0.071 &  0.3929 & 0.000 & -0.2535 & 0.002 \\
			STR × G2                       & -0.2295 & 0.000 & -0.2507 & 0.000 & -0.3051 & 0.000 \\
			STR × G3                       & -0.2907 & 0.000 & -0.2978 & 0.000 & -0.3179 & 0.000 \\
			STR × G4                       & -0.3741 & 0.000 & -0.4974 & 0.000 & -0.3816 & 0.000 \\
			\midrule
			Adjusted $R^2$ & \multicolumn{2}{c}{0.503} & \multicolumn{2}{c}{0.511} & \multicolumn{2}{c}{0.375} \\
			$N$            & \multicolumn{2}{c}{33{,}946} & \multicolumn{2}{c}{33{,}946} & \multicolumn{2}{c}{33{,}946} \\
			\bottomrule
		\end{tabular}
\end{table*}

For \textbf{log(CPM)}, projects in older age groups (G2 to G4) displayed significantly lower baseline commit activity compared to G1 projects, reflecting maturation effects or survivor bias. Among G1 projects, all CE metrics (namely, \texttt{totalIssues\_per\_month}, \texttt{IssueComments\_per\_month}, \texttt{watchers\_per\_month}, and \texttt{stargazers\_per\_month}) positively correlated with commit activity.\\

Crucially, interaction terms revealed age-dependent shifts in these relationships. The influence of \texttt{log(totalIssues\_per\_month)} and \texttt{log(IssueComments\_per\_month)} significantly increased with project age (e.g., IssueComments × G4: 0.4888, $p < 0.001$), underscoring the growing importance of active issue discussion in older projects. In contrast, the predictive power of passive attention metrics (particularly \texttt{log(stargazers\_per\_month)} and, to a lesser extent, \texttt{log(watchers\_per\_month)}) consistently declined with age (e.g., Stargazers × G4: $-0.3741$, $p < 0.001$), suggesting that projects in older groups rely more heavily on core contributors than on broader public visibility.

Patterns for \textbf{log(BPM)} and \textbf{log(RPM)} mirrored those observed for commits. Projects in older groups again showed lower baseline activity. The interaction effects of \texttt{log(IssueComments\_per\_month)} remained consistently positive and significant across all age groups and dynamics. However, \texttt{log(totalIssues\_per\_month)} showed diminishing influence on branches with age and no significant change for releases. Both \texttt{log(watchers\_per\_month)} and \texttt{log(stargazers\_per\_month)} demonstrated weakened correlations with branching and release behavior in older projects.

\begin{minipage}{\linewidth}
	\hrule
	\vspace{5pt}
	\noindent \textbf{Answer to RQ. 2(d):} The influence of individual CE metrics on project dynamics (commits, branches, and releases) varies significantly with project age. Engagement through issue comments becomes increasingly vital for sustaining development activity as projects age, while passive interest (stargazers and watchers) contributes less to core dynamics over time.
	
	\vspace{5pt}
	\hrule
\end{minipage}

\section{Influence of CE and Project Dynamics on Project Lifespan}\label{sec:CEPDLS}

This section assesses the influence of CE and project dynamics on project lifespan, directly addressing RQ~\ref{RQ3}. Many OSS projects face significant survival challenges, with approximately 50\% failing to reach the five-year mark~\cite{ait2022empirical}. For project maintainers, understanding and fostering project longevity is thus critical. Despite its centrality to OSS, community involvement's impact on lifespan remains largely underexplored compared to other dimensions like programming language or license type~\cite{liao2019prediction}. Building on this, we investigate how CE and project dynamic metrics influence project lifespan.

\subsection{Methods}

To evaluate the influence of these metrics, we considered several methodologies, including regression equations and Structural Equation Modeling (SEM). Due to insignificant loadings for lifespan in SEM, and limited variance explainability by regression models, we ultimately opted for a non-parametric statistical approach (Wilcoxon rank-sum test with Cliff's delta) to assess the impact of CE and project dynamics metrics on project lifespan. This decision was primarily driven by the highly skewed and non-normal distributions of activity and engagement rates characteristic of OSS projects, as evidenced in our results (Table~\ref{tab:tab30_combined}). These data characteristics render parametric tests less reliable and make non-parametric methods more robust for comparing distributions and medians across groups.

\subsection{Results}

For a balanced comparison using the Mann-Whitney U test and Cliff's delta, the dataset was partitioned into four lifespan quartiles: short ($<$541 days, Q1), below-average (541 to 1065 days, Q2), above-average (1065 to 1807 days, Q3), and long (1807 to 5360 days, Q4). This quartile division ensures a statistically balanced partitioning of the data, facilitating robust comparisons across distinct lifespan categories and inferring patterns across project lifecycle stages.

Table~\ref{tab:tab30_combined} summarizes the statistical properties of CE and project dynamics metrics across lifespan quartiles. Most metrics show the highest median activity in Q1, followed by a decline with age, suggesting that younger projects attract concentrated early engagement. However, older quartiles exhibit high skewness and kurtosis, indicating that a subset of mature projects continues to receive exceptional attention and development. These patterns underscore CE's role in project longevity, while the nonlinear commit distributions highlight that volume alone may not reflect developmental depth. Factors such as contributor stability and code maturity likely contribute more as projects evolve.

\begin{table}[!h]
	\centering
	\caption{Statistical Properties of CE Metrics and Project Dynamics Across Lifespan Quartiles}
	\label{tab:tab30_combined}
		\begin{tabular}{lcccc}
			\toprule
			\textbf{Metric} & \textbf{Q1} & \textbf{Q2} & \textbf{Q3} & \textbf{Q4} \\
			\midrule
			\multicolumn{5}{c}{\textbf{Community Engagement Metrics}} \\
			\midrule
			\multicolumn{5}{l}{\textbf{TI/m}} \\
			Median    & 1.00 & 0.37 & 0.26 & 0.20 \\
			Kurtosis  & 1893.58 & 674.30 & 144.39 & 314.27 \\
			Skewness  & 35.76 & 21.50 & 9.79 & 13.81 \\
			\midrule
			\multicolumn{5}{l}{\textbf{IC/m}} \\
			Median    & 2.76 & 1.18 & 0.86 & 0.75 \\
			Kurtosis  & 1667.86 & 772.75 & 327.34 & 252.38 \\
			Skewness  & 36.29 & 23.67 & 15.30 & 13.55 \\
			\midrule
			\multicolumn{5}{l}{\textbf{WT/m}} \\
			Median    & 0.62 & 0.22 & 0.15 & 0.12 \\
			Kurtosis  & 1046.75 & 1393.33 & 2336.14 & 122.64 \\
			Skewness  & 29.26 & 27.68 & 37.58 & 8.51 \\
			\midrule
			\multicolumn{5}{l}{\textbf{STR/m}} \\
			Median    & 2.95 & 1.00 & 0.65 & 0.55 \\
			Kurtosis  & 2363.92 & 1074.05 & 398.11 & 226.94 \\
			Skewness  & 41.64 & 27.53 & 16.11 & 12.55 \\
			\midrule
			\multicolumn{5}{c}{\textbf{Project Dynamics Metrics}} \\
			\midrule
			\multicolumn{5}{l}{\textbf{CPM}} \\
			Median    & 10.71 & 3.66 & 2.40 & 1.81 \\
			Kurtosis  & 8001.59 & 2905.46 & 3557.85 & 1906.94 \\
			Skewness  & 88.37 & 50.64 & 57.14 & 42.25 \\
			\midrule
			\multicolumn{5}{l}{\textbf{BPM}} \\
			Median    & 0.28 & 0.09 & 0.05 & 0.03 \\
			Kurtosis  & 4909.47 & 1004.53 & 243.66 & 318.35 \\
			Skewness  & 66.23 & 24.78 & 13.29 & 14.53 \\
			\midrule
			\multicolumn{5}{l}{\textbf{RPM}} \\
			Median    & 0.21 & 0.05 & 0.03 & 0.02 \\
			Kurtosis  & 2930.98 & 945.37 & 625.56 & 850.07 \\
			Skewness  & 51.00 & 22.45 & 19.67 & 22.20 \\
			\bottomrule
		\end{tabular}
\end{table}

To quantify statistical significance and effect sizes, we applied Wilcoxon rank-sum tests and Cliff's delta. Table~\ref{tab:tab34} presents these results.

\begin{table}[!h]
	\centering
	\caption{Pairwise Comparisons Across Lifespan Quartiles (Mann–Whitney U, Cliff's Delta)}
	\label{tab:tab34}
	\setlength{\tabcolsep}{3pt}
	\begin{tabular}{l c c c c c c}
		\toprule
		\multicolumn{7}{c}{\textbf{Quartile 1 vs. Other Lifespan Quartiles}} \\
		\midrule
		& \multicolumn{2}{c}{Q1 vs. Q2} & \multicolumn{2}{c}{Q1 vs. Q3} & \multicolumn{2}{c}{Q1 vs. Q4} \\
		\midrule
		\textbf{Metric} & \textbf{Stat.} & $\delta$ & \textbf{Stat.} & $\delta$ & \textbf{Stat.} & $\delta$ \\
		\midrule
		TI/m   & 49.47M & 0.371 & 53.53M & 0.488 & 55.92M & 0.551 \\
		IC/m   & 46.49M & 0.288 & 49.68M & 0.381 & 51.24M & 0.421 \\
		WT/m   & 53.34M & 0.478 & 58.20M & 0.618 & 60.63M & 0.681 \\
		STR  & 54.62M & 0.514 & 58.69M & 0.631 & 58.93M & 0.634 \\
		CPM   & 51.61M & 0.430 & 56.34M & 0.566 & 59.61M & 0.653 \\
		BPM  & 54.77M & 0.518 & 60.75M & 0.688 & 65.85M & 0.826 \\
		RPM  & 53.46M & 0.482 & 56.84M & 0.580 & 61.70M & 0.711 \\
		\bottomrule
	\end{tabular}
	\vspace{1em}
	\begin{tabular}{l c c c c c c}
		\toprule
		\multicolumn{7}{c}{\textbf{Quartiles 2, 3 vs. Other Lifespan Quartiles}} \\
		\midrule
		& \multicolumn{2}{c}{Q2 vs. Q3} & \multicolumn{2}{c}{Q2 vs. Q4} & \multicolumn{2}{c}{Q3 vs. Q4} \\
		\midrule
		\textbf{Metric} & \textbf{Stat.} & $\delta$ & \textbf{Stat.} & $\delta$ & \textbf{Stat.} & $\delta$ \\
		\midrule
		TI/m   & 41.15M & 0.144 & 44.37M & 0.231 & 39.21M & 0.091 \\
		IC/m   & 39.91M & 0.110 & 41.83M & 0.161 & 37.81M & 0.052 \\
		WT/m   & 43.60M & 0.212 & 47.95M & 0.331 & 40.61M & 0.130 \\
		STR/m  & 45.39M & 0.262 & 47.48M & 0.317 & 40.09M & 0.116 \\
		CPM   & 42.54M & 0.183 & 47.21M & 0.310 & 40.66M & 0.132 \\
		BPM  & 46.23M & 0.286 & 55.72M & 0.546 & 47.33M & 0.317 \\
		RPM  & 48.40M & 0.346 & 53.82M & 0.493 & 49.86M & 0.387 \\
		\bottomrule
	\end{tabular}
\end{table}

The pairwise comparisons of TI/m, IC/m, WT/m, STR/m, and commit-related metrics (CPM, BPM, RPM) across lifespan quartiles (Table~\ref{tab:tab34}) reveal statistically significant differences ($p<0.00104$, Bonferroni corrected), confirming that these metrics vary meaningfully with project lifespan. Cliff’s Delta ($\delta$) values consistently indicate that projects in higher lifespan quartiles generally exhibit greater engagement and activity. Effect sizes increase as the lifespan gap widens, with CE and development metrics showing stronger effects in comparisons between more distant quartiles.

Project Dynamics (CPM, BPM, RPM) exhibit pronounced non-linear trends; effect sizes are moderate in early comparisons but notably stronger for Q1 vs. Q4, suggesting that factors beyond volume (e.g., contributor stability or code maturity) may influence longevity.

Focusing on comparisons between intermediate and older groups, $\delta$ values are generally small to moderate, indicating a gradual increase in engagement and activity. Comparisons such as Q2 vs. Q4 show more noticeable effect sizes, while Q3 vs. Q4 maintains moderate effects, supporting the idea that longer-lived projects retain distinct engagement and development profiles.

Overall, these findings reinforce CE metrics’ role in project longevity, with engagement measures clearly associated with extended lifespans. Commit activity complements this narrative but follows a more nuanced trajectory as projects evolve.

\begin{minipage}[t]{0.95\linewidth}
	\hrule
	\vspace{5pt}
	\noindent \textbf{Answer to RQ. 3:}
	\noindent Yes, per-month rates of CE and project dynamics significantly influence project lifespan, primarily by exhibiting highest activity in younger projects. Longer-lived projects generally show a decline in these per-month rates over time. However, the presence of high skewness and kurtosis in older quartiles indicates that a subset of highly successful and long-lived projects manages to sustain exceptional per-month activity and engagement. This suggests that an initial high burst of activity, followed by either sustained (even if lower) activity or outlier high activity, is associated with project longevity.
	\vspace{5pt}
	\hrule
\end{minipage}

\section{Discussion and Implications}\label{sec:discuss}

This study proposes a novel, comprehensive definition of CE by integrating interdisciplinary community elements within the OSS context. Building on this, we operationalize and validate key engagement metrics, revealing statistically significant associations between CE metrics and both project dynamics and lifespan.

Our central contribution is a CE definition in OSS: an ongoing, interactive process within a Community of Practice (CoP), where diverse members collaboratively pursue shared goals like building high-quality software and fostering sustainability~\cite{Taffere2023,Ye2003}. Emphasizing the CoP aspect and continuous engagement, this definition encompasses all stakeholders (including developers, users, testers, documenters, and supporting organizations)~\cite{MacQueen2001}, offering a holistic view essential for ensuring project longevity~\cite{Hes2019,Kra2019}.

Following this definition, we operationalized CE using measurable metrics derived from ongoing participation and collaborative contributions. These underwent rigorous validation, including Weyuker's properties and face validity checks with OSS users, researchers, and maintainers (see Section~\ref{Metrics Validation}). High agreement affirmed their relevance for analyzing CE's impact.

We identified four GitHub-friendly metrics: Total Issues per Month (TI/m), Issue Comments per Month (IC/m), Watchers per Month (WT/m), and Stargazers per Month (STR/m). TI/m and IC/m reflect problem identification and collaborative discussion; WT/m and STR/m capture broader project interest. This approach aligns with studies on issue tracking~\cite{Zhou2015,Wang2018} and engagement diversity~\cite{Kaur2022}, distinguishing our work from others~\cite{Nijsse2023,Tamburri2019}.

We examined relationships between CE metrics and project dynamics (CPM, BPM, and RPM). Both Active and Passive Engagement Scores were positively associated, especially in highly engaged projects, indicating that vibrant communities contribute more directly to development outcomes. Age-dependent shifts revealed that active engagement grows more influential with maturity, while passive metrics decline (suggesting a shift from public visibility to core collaboration). While insightful, commit-related metrics may not capture documentation or support activity.

We also assessed differences across project lifespans. When normalizing activity by age, younger projects (Q1) showed the highest median per-month rates across all metrics, which declined through Q2–Q4. This emphasizes the formative burst of early engagement as critical for survival. Yet, high skewness and kurtosis in older quartiles (Table~\ref{tab:tab30_combined}) reveal mature projects sustaining exceptional activity as successful outliers. This dual pattern suggests longevity stems from either early intensity or sustained engagement through maturity. 

Implications span multiple stakeholders. For researchers, we offer a validated framework to support consistent longitudinal analysis. For maintainers, fostering early collaboration and sustaining engagement are vital strategies. For the OSS community, our findings empirically link CE to activity and lifespan, illuminating how engagement trajectories shape longevity and suggesting new avenues for sustainability research. Also, these findings, particularly on the evolving influence of engagement across lifecycle stages, are highly generalizable to the distinct phases within continuously active and developing projects.

This study has limitations. Reliability concerns arise from EFA’s subjectivity and latent assumptions. Cliff’s Delta indicates association, not causality. Confounding variables such as popularity, resources, and qualitative aspects challenge interpretation. Additionally, our proxies for dynamics and lifespan definition may limit generalizability. Lastly, Mann–Whitney U assumes independence, so unaddressed project relatedness may affect results. Future work should refine methodologies, broaden metric scope, and account for confounding influences.

\section{Conclusion and Future Work}\label{sec:concl.}
In conclusion, this study provides a comprehensive definition and validated metrics for CE in OSS. We found significant positive associations between CE metrics and project dynamics. The relationship between CE and project lifespan is complex: per-month activity rates are highest in younger projects and generally decline with age, yet a subset of long-lived projects maintains exceptional per-month activity. This suggests an initial burst of CE is vital for project establishment, while sustained high engagement or outlier activity drives long-term longevity. Overall, active CE fosters robust codebases and problem-solving, underscoring its dynamic importance across the OSS lifecycle.

Future work should expand quantitative metrics (e.g., code quality, diversity, user satisfaction) and utilize advanced statistical models (e.g., longitudinal regression, SEM) to control for confounding factors like popularity and age-related factors. Specifically, future longitudinal studies on actively developing projects are crucial to further validate and extend our findings, directly addressing the scope limitation of our current dataset. Incorporating qualitative investigations (e.g., interviews, case studies) and analyzing communication patterns can further elucidate engagement quality. A mixed-methods approach would provide the most comprehensive understanding, while also addressing this study's limitations regarding metric validation, measurement refinement, and controlling other confounding variables.

\section*{Data Availability}
The dataset used in this study is publicly available on Zenodo at \url{https://doi.org/10.5281/zenodo.15244408}.

\bibliographystyle{elsarticle-num}
\bibliography{references2}

\begin{thebibliography}{10}
\expandafter\ifx\csname url\endcsname\relax
  \def\url#1{\texttt{#1}}\fi
\expandafter\ifx\csname urlprefix\endcsname\relax\def\urlprefix{URL }\fi
\expandafter\ifx\csname href\endcsname\relax
  \def\href#1#2{#2} \def\path#1{#1}\fi

\bibitem{Octoverse2022}
\href{https://octoverse.github.com/2022/}{Octoverse 2022: The state of open
  source software}.
\newline\urlprefix\url{https://octoverse.github.com/2022/}

\bibitem{IBMACQ}
\href{https://www.ibm.com/investor/articles/ibm-completes-acquisition-of-red-hat}{Ibm
  completes acquisition of red hat}.
\newline\urlprefix\url{https://www.ibm.com/investor/articles/ibm-completes-acquisition-of-red-hat}

\bibitem{OSSMarket}
\href{https://www.statista.com/statistics/270805/projected-revenue-of-open-source-software-since-2008/}{Open
  source services market revenue worldwide 2017-2022 - statista}.
\newline\urlprefix\url{https://www.statista.com/statistics/270805/projected-revenue-of-open-source-software-since-2008/}

\bibitem{OSSMSize}
\href{https://www.grandviewresearch.com/industry-analysis/open-source-services-market-report}{Open
  source services market size \& share report, 2030}.
\newline\urlprefix\url{https://www.grandviewresearch.com/industry-analysis/open-source-services-market-report}

\bibitem{Schueller2022}
W.~Schueller, J.~Wachs, V.~D. Servedio, S.~Thurner, V.~Loreto,
  \href{https://www.nature.com/articles/s41597-022-01819-z}{Evolving
  collaboration, dependencies, and use in the rust open source software
  ecosystem}, Scientific Data 2022 9:1 9 (2022) 1--10.
\newblock \href {https://doi.org/10.1038/s41597-022-01819-z}
  {\path{doi:10.1038/s41597-022-01819-z}}.
\newline\urlprefix\url{https://www.nature.com/articles/s41597-022-01819-z}

\bibitem{Malgonde2023ResilienceIT}
O.~Malgonde, T.~J.~V. Saldanha, S.~Mithas, Resilience in the open source
  software community: How pandemic and unemployment shocks influence
  contributions to others' and one's own projects, MIS Q. 47 (2023) 361--390.
\newblock \href {https://doi.org/10.25300/misq/2022/17256}
  {\path{doi:10.25300/misq/2022/17256}}.

\bibitem{Gamalielsson2014}
J.~Gamalielsson, B.~Lundell, Sustainability of open source software communities
  beyond a fork: How and why has the libreoffice project evolved?, Journal of
  Systems and Software 89 (2014) 128--145.
\newblock \href {https://doi.org/10.1016/J.JSS.2013.11.1077}
  {\path{doi:10.1016/J.JSS.2013.11.1077}}.

\bibitem{CetrilCommunity}
\href{https://www.cetril.org/community/}{Community in open source software:
  Building connections and collaboration – cetril}.
\newline\urlprefix\url{https://www.cetril.org/community/}

\bibitem{Closson2016}
K.~Closson, R.~McNeil, P.~McDougall, S.~Fernando, A.~B. Collins, R.~B. Turje,
  T.~Howard, S.~Parashar,
  \href{https://pubmed.ncbi.nlm.nih.gov/27717364/}{Meaningful engagement of
  people living with hiv who use drugs: methodology for the design of a peer
  research associate (pra) hiring model}, Harm reduction journal 13 (10 2016).
\newblock \href {https://doi.org/10.1186/S12954-016-0116-Z}
  {\path{doi:10.1186/S12954-016-0116-Z}}.
\newline\urlprefix\url{https://pubmed.ncbi.nlm.nih.gov/27717364/}

\bibitem{Yalegama2016}
S.~Yalegama, N.~Chileshe, T.~Ma, Critical success factors for community-driven
  development projects: A sri lankan community perspective, International
  Journal of Project Management 34 (2016) 643--659.
\newblock \href {https://doi.org/10.1016/J.IJPROMAN.2016.02.006}
  {\path{doi:10.1016/J.IJPROMAN.2016.02.006}}.

\bibitem{Bamberger1987}
M.~Bamberger, The role of community participation in development planning and
  project management (1987).

\bibitem{Bosu2019}
A.~Bosu, K.~Z. Sultana, Diversity and inclusion in open source software (oss)
  projects: Where do we stand?, International Symposium on Empirical Software
  Engineering and Measurement 2019-Septemer (9 2019).
\newblock \href {https://doi.org/10.1109/ESEM.2019.8870179}
  {\path{doi:10.1109/ESEM.2019.8870179}}.

\bibitem{McDonald2013}
N.~McDonald, S.~Goggins, Performance and participation in open source software
  on github, CHI '13 Extended Abstracts on Human Factors in Computing Systems
  2013-April (2013) 139--144.
\newblock \href {https://doi.org/10.1145/2468356.2468382}
  {\path{doi:10.1145/2468356.2468382}}.

\bibitem{Lumbard2024}
K.~Lumbard, M.~Germonprez, S.~Goggins, An empirical investigation of social
  comparison and open source community health, Information Systems Journal 34
  (2024) 499--532.
\newblock \href {https://doi.org/10.1111/ISJ.12485}
  {\path{doi:10.1111/ISJ.12485}}.

\bibitem{West2005}
J.~West, S.~O'Mahony, Contrasting community building in sponsored and community
  founded open source projects, Proceedings of the Annual Hawaii International
  Conference on System Sciences (2005) 196\href
  {https://doi.org/10.1109/HICSS.2005.166} {\path{doi:10.1109/HICSS.2005.166}}.

\bibitem{Hannemann2013}
A.~Hannemann, R.~Klamma, Community dynamics in open source software projects:
  Aging and social reshaping, IFIP Advances in Information and Communication
  Technology 404 (2013) 80--96.
\newblock \href {https://doi.org/10.1007/978-3-642-38928-3_6}
  {\path{doi:10.1007/978-3-642-38928-3_6}}.

\bibitem{MacQueen2001}
K.~M. MacQueen, E.~McLellan, D.~S. Metzger, S.~Kegeles, R.~P. Strauss,
  R.~Scotti, L.~Blanchard, R.~T. Trotter,
  \href{https://pubmed.ncbi.nlm.nih.gov/11726368/}{What is community? an
  evidence-based definition for participatory public health}, American journal
  of public health 91 (2001) 1929--1938.
\newblock \href {https://doi.org/10.2105/AJPH.91.12.1929}
  {\path{doi:10.2105/AJPH.91.12.1929}}.
\newline\urlprefix\url{https://pubmed.ncbi.nlm.nih.gov/11726368/}

\bibitem{Capece2013}
G.~Capece, R.~Costa, The new neighbourhood in the internet era: network
  communities serving local communities, Behavior and Information Technology 32
  (2013) 438--448.
\newblock \href {https://doi.org/10.1080/0144929X.2011.610825}
  {\path{doi:10.1080/0144929X.2011.610825}}.

\bibitem{Bettez2013}
Community building in social justice work: A critical approach, Educational
  Studies 49 (2013) 45--66.
\newblock \href {https://doi.org/10.1080/00131946.2012.749478}
  {\path{doi:10.1080/00131946.2012.749478}}.

\bibitem{Nieckarz2005}
P.~P. Nieckarz, Community in cyber space?: The role of the internet in
  facilitating and maintaining a community of live music collecting and
  trading, City and Community 4 (2005) 403--423.
\newblock \href {https://doi.org/10.1111/j.1540-6040.2005.00145.x}
  {\path{doi:10.1111/j.1540-6040.2005.00145.x}}.

\bibitem{Vogl2017}
G.~Vogl, Work as community: narratives of solidarity and teamwork in the
  contemporary workplace, who owns them?, Sociological Research Online 14~(4)
  (2009) 27--36.
\newblock \href {https://doi.org/10.5153/sro.1933}
  {\path{doi:10.5153/sro.1933}}.

\bibitem{Rothblum2010}
E.~Rothblum, Where is the ‘women’s community?’voices of lesbian,
  bisexual, and queer women and heterosexual sisters, Feminism \& Psychology
  20~(4) (2010) 454--472.
\newblock \href {https://doi.org/10.1177/0959353509355147}
  {\path{doi:10.1177/0959353509355147}}.

\bibitem{Theodori2005}
G.~L. Theodori,
  \href{https://www.tandfonline.com/doi/abs/10.1080/08941920590959640}{Community
  and community development in resource-based areas: Operational definitions
  rooted in an interactional perspective}, Society and Natural Resources 18
  (2005) 661--669.
\newblock \href {https://doi.org/10.1080/08941920590959640}
  {\path{doi:10.1080/08941920590959640}}.
\newline\urlprefix\url{https://www.tandfonline.com/doi/abs/10.1080/08941920590959640}

\bibitem{Brint2001}
S.~Brint, Gemeinschaft revisited: A critique and reconstruction of the
  community concept, Sociological Theory 19 (2001) 1--23.
\newblock \href {https://doi.org/10.1111/0735-2751.00125}
  {\path{doi:10.1111/0735-2751.00125}}.

\bibitem{Cantador2011}
I.~Cantador, P.~Castells, Extracting multilayered communities of interest from
  semantic user profiles: Application to group modeling and hybrid
  recommendations, Computers in Human Behavior 27 (2011) 1321--1336.
\newblock \href {https://doi.org/10.1016/J.CHB.2010.07.027}
  {\path{doi:10.1016/J.CHB.2010.07.027}}.

\bibitem{Theodori2013}
G.~L. Theodori, G.~T. Kyle, Community, place, and conservation, Place-Based
  Conservation: Perspectives from the Social Sciences (2013) 59--70\href
  {https://doi.org/10.1007/978-94-007-5802-5_5}
  {\path{doi:10.1007/978-94-007-5802-5_5}}.

\bibitem{Zacklad2003}
Communities of action (2003) 190--197\href
  {https://doi.org/10.1145/958160.958190} {\path{doi:10.1145/958160.958190}}.

\bibitem{Grout2022}
G.~Grout, What are communities of practice?, Nursing Older People 34 (2022) 15.
\newblock \href {https://doi.org/10.7748/NOP.34.2.15.S6}
  {\path{doi:10.7748/NOP.34.2.15.S6}}.

\bibitem{Ritu2024}
Ritu, A comprehensive study on employee engagement strategies, Educational
  Administration: Theory and Practice 30~(4) (2024) 5894–5899.
\newblock \href {https://doi.org/10.53555/kuey.v30i4.2309}
  {\path{doi:10.53555/kuey.v30i4.2309}}.

\bibitem{Natarajarathinam2021}
M.~Natarajarathinam, S.~Qiu, W.~Lu, Community engagement in engineering
  education: A systematic literature review, Journal of Engineering Education
  110 (2021) 1049--1077.
\newblock \href {https://doi.org/10.1002/JEE.20424}
  {\path{doi:10.1002/JEE.20424}}.

\bibitem{Atlanta1997}
Atlanta, Principles of Community Engagement, 1st Edition, Centres for Disease
  Control and Prevention, 1997.

\bibitem{Born2008}
P.~Born, \href{https://www.paulborn.ca/community-conversations-book}{Community
  conversations : mobilizing the ideas, skills, and passion of community
  organizations, governments, businesses, and people} (2008).
\newline\urlprefix\url{https://www.paulborn.ca/community-conversations-book}

\bibitem{Hes2019}
D.~Hes, C.~Hernandez-Santin, Placemaking fundamentals for the built
  environment, Placemaking Fundamentals for the Built Environment (2019)
  1--326\href {https://doi.org/10.1007/978-981-32-9624-4/COVER}
  {\path{doi:10.1007/978-981-32-9624-4/COVER}}.

\bibitem{Kra2019}
C.~A. Krause-Parello, M.~J. Rice, S.~Sarni, C.~LoFaro, K.~Niitsu,
  M.~McHenry-Edrington, K.~Blanchard, Protective factors for suicide: A
  multi-tiered veteran-driven community engagement project, Journal of Veterans
  Studies 5 (2019) 45.
\newblock \href {https://doi.org/10.21061/JVS.V5I1.111}
  {\path{doi:10.21061/JVS.V5I1.111}}.

\bibitem{Jeyabharathy2023}
D.~P. Jeyabharathy, Aspects and activities of employee engagement,
  INTERANTIONAL JOURNAL OF SCIENTIFIC RESEARCH IN ENGINEERING AND MANAGEMENT 07
  (1 2023).
\newblock \href {https://doi.org/10.55041/IJSREM17478}
  {\path{doi:10.55041/IJSREM17478}}.

\bibitem{hovhannisyan2020theoretical}
A.~Hovhannisyan, V.~Bodrug-Lungu, Theoretical aspects of citizen engagement,
  Studia Universitatis Moldaviae (Seria {\c{S}}tiin{\c{t}}e ale
  Educa{\c{t}}iei) 139~(9) (2020) 135--138.
\newblock \href {https://doi.org/10.5281/zenodo.4277477}
  {\path{doi:10.5281/zenodo.4277477}}.

\bibitem{Taffere2023}
G.~R. Taffere, H.~T. Abebe, Z.~Zerihun, C.~Mallen, H.~P. Price, A.~Mulugeta,
  \href{https://link.springer.com/article/10.1007/s10389-022-01799-9}{Systematic
  review of community engagement approach in research: describing partnership
  approaches, challenges and benefits}, Journal of Public Health (Germany) 32
  (2023) 185--205.
\newblock \href {https://doi.org/10.1007/S10389-022-01799-9/TABLES/2}
  {\path{doi:10.1007/S10389-022-01799-9/TABLES/2}}.
\newline\urlprefix\url{https://link.springer.com/article/10.1007/s10389-022-01799-9}

\bibitem{Ye2003}
Y.~Ye, K.~Kishida, Toward an understanding of the motivation of open source
  software developers, Proceedings - International Conference on Software
  Engineering (2003) 419--429\href {https://doi.org/10.1109/ICSE.2003.1201220}
  {\path{doi:10.1109/ICSE.2003.1201220}}.

\bibitem{Tamburri2019}
D.~A. Tamburri, F.~Palomba, A.~Serebrenik, A.~Zaidman, Discovering community
  patterns in open-source: a systematic approach and its evaluation, Empirical
  Software Engineering 24 (2019) 1369--1417.
\newblock \href {https://doi.org/10.1007/S10664-018-9659-9/FIGURES/8}
  {\path{doi:10.1007/S10664-018-9659-9/FIGURES/8}}.

\bibitem{Daniel2012}
S.~Daniel, R.~Agarwal, K.~J. Stewart, The effects of diversity in global,
  distributed collectives: A study of open source project success, Information
  systems research 24~(2) (2013) 312--333.
\newblock \href {https://doi.org/10.1287/isre.1120.0435}
  {\path{doi:10.1287/isre.1120.0435}}.

\bibitem{Chodapaneedi2017}
M.~T. Chodapaneedi, S.~Manda,
  \href{https://urn.kb.se/resolve?urn=urn:nbn:se:bth-15431}{Engagement of
  developers in open source projects : A multi-case study} (2017).
\newline\urlprefix\url{https://urn.kb.se/resolve?urn=urn:nbn:se:bth-15431}

\bibitem{Nijsse2023}
J.~Nijsse, A.~Litchfield, Identifying developer engagement in open-source
  software blockchain projects through factor analysis (1 2023).
\newblock \href {https://doi.org/10.24251/HICSS.2023.651}
  {\path{doi:10.24251/HICSS.2023.651}}.

\bibitem{Nijsem2023}
J.~Nijsse, A.~Litchfield, \href{https://arxiv.org/abs/2310.20277v1}{Towards a
  structural equation model of open source blockchain software health} (10
  2023).
\newline\urlprefix\url{https://arxiv.org/abs/2310.20277v1}

\bibitem{Stewart2006}
K.~J. Stewart, S.~Gosain, The impact of ideology on effectiveness in open
  source software development teams, MIS Quarterly: Management Information
  Systems 30 (2006) 291--314.
\newblock \href {https://doi.org/10.2307/25148732}
  {\path{doi:10.2307/25148732}}.

\bibitem{Zhou2015}
M.~Zhou, A.~Mockus, Who will stay in the floss community? modeling
  participant's initial behavior, IEEE Transactions on Software Engineering 41
  (2015) 82--99.
\newblock \href {https://doi.org/10.1109/TSE.2014.2349496}
  {\path{doi:10.1109/TSE.2014.2349496}}.

\bibitem{Wang2018}
T.~Wang, Y.~Zhang, G.~Yin, Y.~Yu, H.~Wang, Who will become a long-term
  contributor? a prediction model based on the early phase behaviors, ACM
  International Conference Proceeding Series (9 2018).
\newblock \href {https://doi.org/10.1145/3275219.3275223}
  {\path{doi:10.1145/3275219.3275223}}.

\bibitem{Constantinou2017}
E.~Constantinou, T.~Mens, An empirical comparison of developer retention in the
  rubygems and npm software ecosystems, Innovations in Systems and Software
  Engineering 13 (2017) 101--115.
\newblock \href {https://doi.org/10.1007/S11334-017-0303-4/FIGURES/10}
  {\path{doi:10.1007/S11334-017-0303-4/FIGURES/10}}.

\bibitem{Kaur2022}
R.~Kaur, K.~K. Chahal, M.~Saini, Understanding community participation and
  engagement in open source software projects: A systematic mapping study,
  Journal of King Saud University - Computer and Information Sciences 34 (2022)
  4607--4625.
\newblock \href {https://doi.org/10.1016/J.JKSUCI.2020.10.020}
  {\path{doi:10.1016/J.JKSUCI.2020.10.020}}.

\bibitem{Norikane2017}
T.~Norikane, A.~Ihara, K.~Matsumoto, Which review feedback did long-term
  contributors get on oss projects?, SANER 2017 - 24th IEEE International
  Conference on Software Analysis, Evolution, and Reengineering (2017)
  571--572\href {https://doi.org/10.1109/SANER.2017.7884682}
  {\path{doi:10.1109/SANER.2017.7884682}}.

\bibitem{duenas2018perceval}
S.~Due{\~n}as, V.~Cosentino, G.~Robles, J.~M. Gonzalez-Barahona, Perceval:
  software project data at your will, in: Proceedings of the 40th international
  conference on software engineering: companion proceeedings, 2018, pp. 1--4.

\bibitem{goggins2021making}
S.~P. Goggins, M.~Germonprez, K.~Lumbard, Making open source project health
  transparent, Computer 54~(8) (2021) 104--111.

\bibitem{franch2014layered}
X.~Franch, R.~Kenett, F.~Mancinelli, A.~Susi, D.~Ameller, R.~Ben-Jacob,
  A.~Siena, A layered approach to managing risks in oss projects, in: IFIP
  International Conference on Open Source Systems, Springer, 2014, pp.
  168--171.

\bibitem{noda2023devex}
A.~Noda, M.-A. Storey, N.~Forsgren, M.~Greiler, Devex: What actually drives
  productivity: The developer-centric approach to measuring and improving
  productivity, Queue 21~(2) (2023) 35--53.

\bibitem{Dabic2021}
O.~Dabic, E.~Aghajani, G.~Bavota, Sampling projects in github for msr studies,
  Proceedings - 2021 IEEE/ACM 18th International Conference on Mining Software
  Repositories, MSR 2021 (2021) 560--564\href
  {https://doi.org/10.1109/MSR52588.2021.00074}
  {\path{doi:10.1109/MSR52588.2021.00074}}.

\bibitem{Hata2022}
H.~Hata, N.~Novielli, S.~Baltes, R.~G. Kula, C.~Treude, Github discussions: An
  exploratory study of early adoption, Empirical Software Engineering 27 (2022)
  1--32.
\newblock \href {https://doi.org/10.1007/S10664-021-10058-6/FIGURES/4}
  {\path{doi:10.1007/S10664-021-10058-6/FIGURES/4}}.

\bibitem{liao2019prediction}
Z.~Liao, B.~Zhao, S.~Liu, H.~Jin, D.~He, L.~Yang, Y.~Zhang, J.~Wu, A prediction
  model of the project life-span in open source software ecosystem, Mobile
  Networks and Applications 24 (2019) 1382--1391.
\newblock \href {https://doi.org/10.1007/s11036-018-0993-3}
  {\path{doi:10.1007/s11036-018-0993-3}}.

\bibitem{xia2022predicting}
T.~Xia, W.~Fu, R.~Shu, R.~Agrawal, T.~Menzies, Predicting health indicators for
  open source projects (using hyperparameter optimization), Empirical Software
  Engineering 27~(6) (2022) 122.
\newblock \href {https://doi.org/10.1007/s10664-022-10171-0}
  {\path{doi:10.1007/s10664-022-10171-0}}.

\bibitem{Zhang2022}
Building a community of practice in the workplace, International Journal of
  Smart Education and Urban Society 13 (2022) 1--11.
\newblock \href {https://doi.org/10.4018/IJSEUS.304362}
  {\path{doi:10.4018/IJSEUS.304362}}.

\bibitem{santos2018communities}
L.~P. Santos, L.~N. Barbosa, D.~A. Bessa, L.~P. Martins, L.~S. Barbosa,
  Communities of practice as a tool to support the gcio function, in:
  Proceedings of the 11th International Conference on Theory and Practice of
  Electronic Governance, 2018, pp. 118--126.
\newblock \href {https://doi.org/10.1145/3209415.32095}
  {\path{doi:10.1145/3209415.32095}}.

\bibitem{Langley2017}
A.~Langley, H.~Patel, R.~J. Houghton, Fostering a community of practice for
  industrial processes, Dynamics of Long-Life Assets: From Technology
  Adaptation to Upgrading the Business Model (2017) 151--168\href
  {https://doi.org/10.1007/978-3-319-45438-2_9/COVER}
  {\path{doi:10.1007/978-3-319-45438-2_9/COVER}}.

\bibitem{moradi2021community}
B.~Moradi-Jamei, B.~L. Kramer, J.~B.~S. Calder{\'o}n, G.~Korkmaz, Community
  formation and detection on github collaboration networks, in: Proceedings of
  the 2021 IEEE/ACM International Conference on Advances in Social Networks
  Analysis and Mining, 2021, pp. 244--251.
\newblock \href {https://doi.org/10.1145/3487351.3488278}
  {\path{doi:10.1145/3487351.3488278}}.

\bibitem{Han2023}
Y.~Han, Z.~Wang, Y.~Feng, Z.~Zhao, Y.~Wang, Cross-status communication and
  project outcomes in oss development: A language style matching perspective,
  Empirical Software Engineering 28 (2023) 1--36.
\newblock \href {https://doi.org/10.1007/S10664-023-10298-8/TABLES/12}
  {\path{doi:10.1007/S10664-023-10298-8/TABLES/12}}.

\bibitem{liang2022understanding}
J.~T. Liang, T.~Zimmermann, D.~Ford, Understanding skills for oss communities
  on github, in: Proceedings of the 30th ACM Joint European Software
  Engineering Conference and Symposium on the Foundations of Software
  Engineering, 2022, pp. 170--182.
\newblock \href {https://doi.org/10.1145/3540250.3549082}
  {\path{doi:10.1145/3540250.3549082}}.

\bibitem{Hanna2018}
H.~Mäenpää, S.~Mäkinen, T.~Kilamo, T.~Mikkonen, T.~Männistö, P.~Ritala,
  Organizing for openness: six models for developer involvement in hybrid oss
  projects, Journal of Internet Services and Applications 9 (2018) 1--14.
\newblock \href {https://doi.org/10.1186/S13174-018-0088-1/TABLES/3}
  {\path{doi:10.1186/S13174-018-0088-1/TABLES/3}}.

\bibitem{Trinkenreich2020}
B.~Trinkenreich, M.~Guizani, I.~Wiese, A.~Sarma, I.~Steinmacher, Hidden
  figures: Roles and pathways of successful oss contributors, Proceedings of
  the ACM on Human-Computer Interaction 4 (2020) 22.
\newblock \href {https://doi.org/10.1145/3415251} {\path{doi:10.1145/3415251}}.

\bibitem{awe2022comprehensive}
O.~O. Awe, P.~O. Jegede, J.~Cochran, A comprehensive tutorial on factor
  analysis with r: Empirical insights from an educational perspective,
  Promoting statistical practice and collaboration in developing countries 265
  (2022) 9781003261148--24.

\bibitem{Kim2019}
J.~H. Kim, Multicollinearity and misleading statistical results, Korean Journal
  of Anesthesiology 72 (2019) 558.
\newblock \href {https://doi.org/10.4097/KJA.19087}
  {\path{doi:10.4097/KJA.19087}}.

\bibitem{Sheoran2014}
J.~Sheoran, K.~Blincoe, E.~Kalliamvakou, D.~Damian, J.~Ell, Understanding
  "watchers" on github, 11th Working Conference on Mining Software
  Repositories, MSR 2014 - Proceedings (2014) 336--339\href
  {https://doi.org/10.1145/2597073.2597114}
  {\path{doi:10.1145/2597073.2597114}}.

\bibitem{bertram2010communication}
D.~Bertram, A.~Voida, S.~Greenberg, R.~Walker, Communication, collaboration,
  and bugs: the social nature of issue tracking in small, collocated teams, in:
  Proceedings of the 2010 ACM conference on Computer supported cooperative
  work, 2010, pp. 291--300.
\newblock \href {https://doi.org/10.1145/1718918.1718972}
  {\path{doi:10.1145/1718918.1718972}}.

\bibitem{Xia2019}
Y.~Xia, Y.~Yang, Rmsea, cfi, and tli in structural equation modeling with
  ordered categorical data: The story they tell depends on the estimation
  methods, Behavior Research Methods 51 (2019) 409--428.
\newblock \href {https://doi.org/10.3758/S13428-018-1055-2/TABLES/9}
  {\path{doi:10.3758/S13428-018-1055-2/TABLES/9}}.

\bibitem{rath2020request}
M.~Rath, P.~M{\"a}der, Request for comments: conversation patterns in issue
  tracking systems of open-source projects, in: Proceedings of the 35th Annual
  ACM Symposium on Applied Computing, 2020, pp. 1414--1417.
\newblock \href {https://doi.org/10.1145/3341105.3374056}
  {\path{doi:10.1145/3341105.3374056}}.

\bibitem{srinivasan2014software}
K.~Srinivasan, T.~Devi, Software metrics validation methodologies in software
  engineering, International Journal of Software Engineering \& Applications
  5~(6) (2014) 87.

\bibitem{misra2008applicability}
S.~Misra, I.~Akman, Applicability of weyuker's properties on oo metrics: Some
  misunderstandings, Computer Science and Information Systems 5~(1) (2008)
  17--23.
\newblock \href {https://doi.org/10.2298/CSIS0801017M}
  {\path{doi:10.2298/CSIS0801017M}}.

\bibitem{de2013applicability}
D.~De~Silva, N.~Kodagoda, Applicability of weyuker's properties using three
  complexity metrics, in: 2013 8th International Conference on Computer Science
  \& Education, IEEE, 2013, pp. 685--690.
\newblock \href {https://doi.org/10.1109/ICCSE.2013.6553996}
  {\path{doi:10.1109/ICCSE.2013.6553996}}.

\bibitem{Saraf2013}
N.~Saraf, A.~Seary, D.~Chandrasekaran, P.~Monge, Evolution of an open source
  community network – an exploratory study, SSRN Electronic Journal (2 2013).
\newblock \href {https://doi.org/10.2139/SSRN.2005302}
  {\path{doi:10.2139/SSRN.2005302}}.

\bibitem{jarczyk2014github}
O.~Jarczyk, B.~Gruszka, S.~Jaroszewicz, L.~Bukowski, A.~Wierzbicki, Github
  projects. quality analysis of open-source software, in: Social Informatics:
  6th International Conference, SocInfo 2014, Barcelona, Spain, November 11-13,
  2014. Proceedings 6, Springer, 2014, pp. 80--94.
\newblock \href {https://doi.org/10.1007/978-3-319-13734-6_6}
  {\path{doi:10.1007/978-3-319-13734-6_6}}.

\bibitem{ait2022empirical}
A.~Ait, J.~L.~C. Izquierdo, J.~Cabot, An empirical study on the survival rate
  of github projects, in: Proceedings of the 19th International Conference on
  Mining Software Repositories, 2022, pp. 365--375.
\newblock \href {https://doi.org/10.1145/3524842.3527941}
  {\path{doi:10.1145/3524842.3527941}}.

\end{thebibliography}
\end{document}